\documentclass[a4paper,fleqn,usenatbib]{mnras}

\usepackage[T1]{fontenc}
\usepackage{ae,aecompl}

\newcommand{\hl}[1]{{#1}}  %
\usepackage{graphicx}   
\usepackage{amsmath}    
\usepackage{bm}       
\usepackage{amssymb}    
\usepackage{wasysym}  
\usepackage{txfonts}
\usepackage{etoolbox}
\makeatletter 
  \patchcmd{\NAT@citex}
    {\@citea\NAT@hyper@{%
      \NAT@nmfmt{\NAT@nm}%
      \hyper@natlinkbreak{\NAT@aysep\NAT@spacechar}{\@citeb\@extra@b@citeb}%
      \NAT@date}}
    {\@citea\NAT@nmfmt{\NAT@nm}%
    \NAT@aysep\NAT@spacechar\NAT@hyper@{\NAT@date}}{}{}

  \patchcmd{\NAT@citex}
    {\@citea\NAT@hyper@{%
      \NAT@nmfmt{\NAT@nm}%
      \hyper@natlinkbreak{\NAT@spacechar\NAT@@open\if*#1*\else#1\NAT@spacechar\fi}%
        {\@citeb\@extra@b@citeb}%
      \NAT@date}}
    {\@citea\NAT@nmfmt{\NAT@nm}%
    \NAT@spacechar\NAT@@open\if*#1*\else#1\NAT@spacechar\fi\NAT@hyper@{\NAT@date}}
    {}{}
\makeatother

\newcommand\Msun{\text{M}_{\astrosun}} 
\newcommand\Lsun{\text{L}_{\astrosun}} 
\newcommand\Zsun{\text{Z}_{\astrosun}} 
\newcommand\colt{\text{COLT}} 
\newcommand\HI{{H\,\textsc{i}}} 
\newcommand\HII{{H\,\textsc{ii}}} 
\newcommand\HeI{{He\,\textsc{i}}} 
\newcommand\HeII{{He\,\textsc{ii}}} 

\title[Ly$\alpha$ radiation hydrodynamics of galactic winds before reionization]{Lyman $\bm{\alpha}$ radiation hydrodynamics of galactic winds \\ before cosmic reionization}

\author[A.\ Smith et al.]{
  Aaron~Smith,$^1$\thanks{E-mail: \href{mailto:asmith@astro.as.utexas.edu}{asmith@astro.as.utexas.edu}}
  Volker~Bromm$^1$ and
  Abraham~Loeb$^2$
  \\
  $^1$Department of Astronomy, The University of Texas at Austin, Austin, TX 78712, USA \\
  $^2$Department of Astronomy, Harvard University, 60 Garden Street, Cambridge, MA 02138, USA
}

\date{Accepted XXX. Received YYY; in original form ZZZ}

\pubyear{2016}

\begin{document}
\label{firstpage}
\pagerange{\pageref{firstpage}--\pageref{lastpage}}
\maketitle

\begin{abstract}
  The dynamical impact of Lyman~$\alpha$~(Ly$\alpha$) radiation pressure on galaxy formation depends on the rate and duration of momentum transfer between Ly$\alpha$ photons and neutral hydrogen gas. Although photon trapping has the potential to multiply the effective force, ionizing radiation from stellar sources may relieve the Ly$\alpha$ pressure before appreciably affecting the kinematics of the host galaxy or efficiently coupling Ly$\alpha$ photons to the outflow. We present self-consistent Ly$\alpha$ radiation-hydrodynamics simulations of high-$z$ galaxy environments by coupling the Cosmic Ly$\alpha$ Transfer code ({\sc colt}) with spherically symmetric Lagrangian frame hydrodynamics. The accurate but computationally expensive Monte-Carlo radiative transfer calculations are feasible under the one-dimensional approximation. \hl{The initial starburst drives an expanding shell of gas from the centre and} in certain cases Ly$\alpha$ feedback significantly enhances the \hl{shell velocity}. Radiative feedback alone is capable of ejecting baryons into the intergalactic medium~(IGM) for protogalaxies with a virial mass of $M_{\rm vir} \lesssim 10^8~\Msun$. We compare the Ly$\alpha$ signatures of Population~III stars with $10^5$~K blackbody emission to that of direct collapse black holes with a nonthermal Compton-thick spectrum and find substantial differences \hl{if the Ly$\alpha$ spectra are shaped by gas pushed by Ly$\alpha$ radiation-driven winds}. For both sources, the flux emerging from the galaxy is reprocessed by the IGM such that the observed Ly$\alpha$ luminosity is reduced significantly and the time-averaged velocity offset of the Ly$\alpha$ peak is shifted redward.
\end{abstract}



\section{Introduction}
\label{sec:introduction}
Radiation from the first stars and galaxies initiated a dramatic transformation throughout the Universe, marking the end of the cosmic dark ages \citep{Bromm_Yoshida_2011,Loeb_Furlanetto_2013}. The observational frontier for high-redshift galaxies has been extended into the epoch of reionization \citep{Bouwens_2011,Finkelstein_2013,Oesch_van_Dokkum_2015,Stark_2015,Zitrin_2015,Oesch_2016}. Furthermore, next-generation observatories such as the \textit{James Webb Space Telescope}~\citep[\textit{JWST};][]{Gardner_2006} will probe even deeper into the past and provide essential details about cosmic history. The Ly$\alpha$ transition of neutral hydrogen (\HI) plays a prominent role in spectral observations of high-$z$ objects. However, due to the high opacity of pre-reionized gas, direct detection is challenging. Still, it may be possible to observe the indirect signatures of radiatively-driven outflows, including from Ly$\alpha$ radiation pressure which is more prominent in these conditions.

Within the first galaxies, up to two-thirds of the ionizing photons from massive stars are reprocessed into Ly$\alpha$ radiation \citep{Partridge_Peebles_1967,Dijkstra_2014}. However, because neutral hydrogen is opaque to the Ly$\alpha$ line, photon trapping effectively acts as a force multiplier applied to gas surrounding \HII\ regions. Indeed, the role of Ly$\alpha$ radiation pressure throughout the galactic assembly process has been discussed extensively, particularly in the context of other feedback mechanisms \citep[e.g.][]{Cox_1985,Haehnelt_1995,Oh_Haiman_2002,McKee_Tan_2008}. Such discussions tend to focus on order of magnitude estimates based on idealized radiative transfer calculations \citep{Wise_2012,Dijkstra_Loeb_2008,Milosavljevic_2009}. Up to this point, accurate radiation-hydrodynamics (RHD) simulations incorporating Ly$\alpha$ feedback have not been performed, either because the effects are considered sub-dominant or the perceived computational costs prohibited such a treatment. However, the nature of this question requires full consideration of the dynamical coupling between matter and radiation.

The first galaxies were likely atomic cooling haloes whose virial temperatures activate Ly$\alpha$ line cooling, i.e. $T_\text{vir} \gtrsim 10^4$\,K \citep{Bromm_Yoshida_2011}. In this framework, the first galaxies greatly impacted their surroundings as the initial drivers of reionization. Furthermore, radiative feedback from Population~III~or~II stars dramatically altered the gas within these relatively low-mass systems. Feedback physics is crucial for understanding the multi-scale connections of astrophysical phenomena and their observational signatures. So far simulations have focused on thermally-driven supernova (SN) feedback while the impact of Ly$\alpha$ radiation pressure is still relatively unexplored.

The physical processes that couple Ly$\alpha$ radiation to gas dynamics are continuum absorption by dust and momentum transfer via multiple scattering with neutral hydrogen. These two mechanisms are roughly independent of each other in the sense that high dust content reduces the Ly$\alpha$ escape fraction while an absence of dust results in a pure scattering scenario. \citet{Wise_2012} argue for the existence of a metallicity upper limit such that if $Z \gtrsim 0.05~\Zsun$ then Ly$\alpha$ radiation pressure may be ignored \hl{because of the increasing impact of dust opacity} \citep[see also][]{Henney_1998}. \hl{However,} it is uncertain whether metallicity thresholds are universally applicable considering the nontrivial nature of Ly$\alpha$ radiative transfer in inhomogeneous, dusty media. Throughout this paper we approximate first galaxies as metal-free environments so dust effects are not discussed in detail. Nonetheless, even without absorption, Ly$\alpha$ photon trapping only affects the residual \HI\ within ionized regions. Therefore, unless the gas remains neutral for a long enough duration even relatively strong sources are kinematically inconsequential. Furthermore, geometric effects such as gas clumping, rotation, and filamentary structure often lead to anisotropic escape, photon leakage, or otherwise altered dynamical impact.

In regions dominated by Ly$\alpha$ radiation pressure we expect expansion to set in and eventually limit the impact of subsequent feedback. Historically, this was recognized to be important in the context of planetary nebulae as far back as \citet{Ambarzumian_1932}, \citet{Zanstra_1934}, \citet{Struve_1942}, and \citet{Chandrasekhar_1945}. The authors found that expansion could lower the Ly$\alpha$ opacity to the extent that the radiation pressures due to Ly$\alpha$ and Lyman continuum are the same order of magnitude. Later, \citet{Cox_1985} examined Ly$\alpha$ pressure in the context of providing disc support during the formation epoch of spiral galaxies. \citet{Bithell_1990} and \citet{Haehnelt_1995} claimed further importance in galactic modeling by arguing that fully ionized, self-gravitating objects can be supported by radiation pressure for characteristic lengths of $\ell_\alpha \sim 100~\text{pc} - 3~\text{kpc}$. More recently, \citet{Dijkstra_Loeb_2008} found that multiple scattering within high \HI\ column density shells is capable of enhancing the effective Ly$\alpha$ radiation pressure by one or two orders of magnitude. The static case sets the upper limit on force multiplication while subsequent acceleration renders an ever diminishing effective opacity \citep{Dijkstra_Loeb_2009}. The exact nature of the dynamics of Ly$\alpha$-driven winds and the extent of their impact is an intriguing yet challenging problem.

Still, there are other scenarios in which Ly$\alpha$ feedback may play a significant role. For example, \citet{Oh_Haiman_2002} argue that Ly$\alpha$ trapping can constrain the efficiency of star formation in high-$z$ galaxies. This follows from a discussion by \citet{Rees_Ostriker_1977} in which cooling radiation attempts to unbind a system in free-fall collapse. In this picture, although radiation pressure does not overcome collapse it increases the temperature and the boosted Jeans mass is likely to inhibit fragmentation \citep[see also][]{Latif_2011}. \citet{Oh_Haiman_2002} conclude that Ly$\alpha$ photon pressure is likely to be an important source of feedback until supernova explosions become dominant. In fact, Ly$\alpha$ radiation pressure may contribute alongside other feedback mechanisms in the formation of intermediate-mass ``seed'' black holes \citep{Dijkstra_Haiman_2008,Milosavljevic_2009,Smith_CR7_2016}. Finally, Ly$\alpha$ radiation pressure has also been studied in the context of the first stars, where it may reverse outflow along the polar directions but is not likely to be significant elsewhere \citep{McKee_Tan_2008,Stacy_2012}.

Much of the uncertainty regarding Ly$\alpha$ dynamics is related to difficulties in numerical modeling. Monte-Carlo radiative transfer (MCRT) has emerged as the prevalent method for accurate Ly$\alpha$ calculations \citep{Ahn_2002,Zheng_2002,Dijkstra_2006}. In many cases Ly$\alpha$ transfer codes are used to post-process realistic hydrodynamical simulations \citep[e.g.][]{Tasitsiomi_2006,Laursen_2009,Verhamme_2012,Smith_2015}. However, idealized models described by a few basic parameters have also been widely used to study Ly$\alpha$ spectra from moderate redshift galaxies \citep{Ahn_2004,Verhamme_2006,Gronke_2015,Gronke_Multiphase_2016}. Along these lines, \citet{Dijkstra_Loeb_2008} performed the first, direct MCRT calculations of Ly$\alpha$ radiation pressure for various spherically symmetric models representing different stages of galaxy formation. The authors describe scenarios for supersonic Ly$\alpha$-driven outflows for $\sim 10^6~\Msun$ minihalo environments. They argue that Ly$\alpha$ pressure is too weak to affect larger ($\gtrsim 10^9~\Msun$) high-redshift star-forming galaxies, except in the case of $\lesssim 1$~kpc galactic supershells in the interstellar medium as explored in greater detail by \citet{Dijkstra_Loeb_2009}. We emphasize that each of the above estimates are based on non-dynamical simulations. Still, we must consider the possibility that Ly$\alpha$ radiation can drive galactic winds and affect regions with neutral gas throughout the galaxy formation process. In contrast to the MCRT method, \citet{Latif_2011} use the stiffened equation of state proposed by \citet{Spaans_Silk_2006} to mimic Ly$\alpha$ pressure effects in a cosmological simulation. However, it is unclear whether such an approach faithfully reproduces the complex physics involved with Ly$\alpha$ radiative transfer. The qualitative result is similar to other forms of radiation pressure, which collectively regulate star formation by driving turbulence and increase the efficiency of supernova-driven outflows \citep{Wise_2012}.

A substantial effort has recently been invested in RHD simulations using MCRT, which is often more accurate than other methods but comes at a higher computational cost \citep[e.g.][]{Abdikamalov_2012,Harries_2015,Roth_2015,Tsang_2015}. Indeed, a dynamical approach seems timely for Ly$\alpha$ feedback because of increasingly powerful computational resources in conjunction with improved theoretical and numerical algorithms \citep[e.g.][]{Mihalas_book_1984,Mihalas_2001,Castor_book_2004}. Radiation hydrodynamics also provides a more general context for Ly$\alpha$ studies than allowed from post-processing simulations. The dynamical context may also provide new information about the likelihood of observations at various stages of galaxy evolution, especially before and after reionization. Cosmological RHD simulations have undergone significant advances in recent years \citep{Jeon_2015,Norman_2015,So_2015}. There are hints that radiative feedback will boost the escape fraction of ionizing radiation in the shallow potential wells of the first galaxies, which may significantly impact their visibility \citep{Pawlik_2013}. These lower mass systems may quickly respond to Ly$\alpha$ radiation, ionizing radiation, supernova explosions, and other feedback mechanisms which modify aspects of the standard picture of galaxy and star formation.

The observed Ly$\alpha$ flux depends on properties of the host galaxy, the intervening IGM, and the external sight line to the detector. \hl{Previous studies have shown that asymmetries in the gas distribution introduce a directional dependence of the spectrum, which can be strongly enhanced by the presence of significantly lower-column density pathways as is the case for the models of} \citet{Behrens_2014} and \citet{Dijkstra_DCBH_2016}. Still, the intrinsic escape of Ly$\alpha$ photons from individual galaxies likely deviates from isotropy by at most a factor of a few, \hl{especially when considering Ly$\alpha$ radiative transfer effects due to the larger-scale environment} \citep{Smith_2015}. Therefore, we focus our study of Ly$\alpha$ feedback on spherically symmetric setups, which \hl{affects the robustness of observational predictions but} is also more computationally feasible because the aggregate statistics converge with fewer photon packets. This paper is organized as follows. In Section~\ref{sec:cosmological_context} we discuss the cosmological context and provide analytic estimates for Ly$\alpha$ feedback effects. In Section~\ref{sec:radiative_transfer_methodology} we present the radiative transfer methodology to obtain Monte-Carlo estimates of local dynamical quantities, including tests of our code against known solutions. In Section~\ref{sec:radiation_hydrodynamics_methodology} we present the remaining hydrodynamics methodology along with additional physics required for our models. In Section~\ref{sec:dynamical_impact} we present Ly$\alpha$ RHD simulations of galaxies with different strengths for the central starburst or black hole. This is intended to determine the dynamical impact of Ly$\alpha$ radiation pressure during the early stages of galaxy formation. Finally, in Section~\ref{sec:conc} we conclude.

\section{The Cosmological Context}
\label{sec:cosmological_context}
Ly$\alpha$ radiation pressure from the first stars likely had a substantial impact on their host environments. Relatively low mass minihaloes with $M_\text{vir} \lesssim 10^6~\Msun$ would have been especially susceptible to Ly$\alpha$ scattering against the neutral gas surrounding the central \HII\ region. A similar scenario is also possible for \HI\ ``supershells'' in interstellar environments within more massive galaxies later in cosmic history. However, the overall impact of Ly$\alpha$ trapping diminishes rapidly as ongoing star formation and AGN activity enlarge the ionized bubbles and eventually reionize the Universe. In the remainder of this section we present analytical estimates of Ly$\alpha$ feedback effects to determine the relative importance in various contexts. In some cases these already serve as test cases of our code.

\subsection{Sources in a neutral expanding IGM}
\label{sub:sources_in_a_neutral_comoving_igm}
We now consider the radiation pressure due to a Ly$\alpha$ point source at redshift $z = 10$ embedded in a neutral, homogeneous intergalactic medium undergoing Hubble expansion. As the photons scatter and diffuse they eventually experience enough cosmological redshifting that they move far into the red wing of the line profile and free streaming becomes unavoidable. \citet{Loeb_Rybicki_1999} calculated an analytic solution for the angle-averaged intensity~$J(\nu, r)$ as a function of radius valid in the diffusion limit. This is an ideal test case because the initial setup is clear, analytic expressions for Ly$\alpha$ radiation quantities are readily available, and we may compare to previous results obtained by \citet{Dijkstra_Loeb_2008}.

We briefly describe our setup parameters. The neutral hydrogen background number density is $n_\text{H, IGM} \approx 2.5 \times 10^{-4}$~cm$^{-3}$ $[(1+z)/11]^3$ throughout the entire domain. The solution from \citet{Loeb_Rybicki_1999} assumes a zero temperature limit despite the higher cosmic background temperature of $T_\text{CMB} \approx 30$~K. Therefore, in order to minimize thermal effects we use a uniform temperature of $T = 1$~K. Our grid consists of concentric spherical shells spaced such that the thickness increases with radius according to $\Delta r \propto r^{1/2}$. The minimum radius $r_\text{min} \approx 50$~pc is chosen to achieve a reasonable resolution up to the maximum radius $r_\text{max} \approx 10$~Mpc where the IGM is optically thin to redshifted Ly$\alpha$ photons. The velocity field follows from an isotropic expansion law of $v(r) = H(z) r$, where the redshift-dependent Hubble parameter is $H(z) \approx H_0 \Omega_\text{m}^{1/2} (z+1)^{3/2}$. Here we assume a flat matter-dominated high-$z$ Universe with a present-day Hubble constant of $H_0 = 67.8$~km~s$^{-1}$~Mpc$^{-1}$ and matter density parameter of $\Omega_\text{m} \approx 0.3$. Specifically, in terms of the characteristic scales discussed by \citet{Loeb_Rybicki_1999} the comoving frequency shift from line centre at which the optical depth reaches unity is $\nu_\ast \approx 1.2 \times 10^{13}$~Hz~$[(1+z)/11]^{3/2}$ so that $\tau_\text{IGM}(\nu_\ast) \equiv 1$. This corresponds to a proper radius $r_\ast \approx 1$~Mpc for which the Doppler shift due to Hubble expansion produces the critical frequency shift $\nu_\ast$. Finally, the relative proper velocity at the critical radius is $v_\ast \equiv H(z) r_\ast |_{z=10} \approx 1450$~km~s$^{-1}$.

  \begin{figure}
    \centering
    \includegraphics[width=\columnwidth]{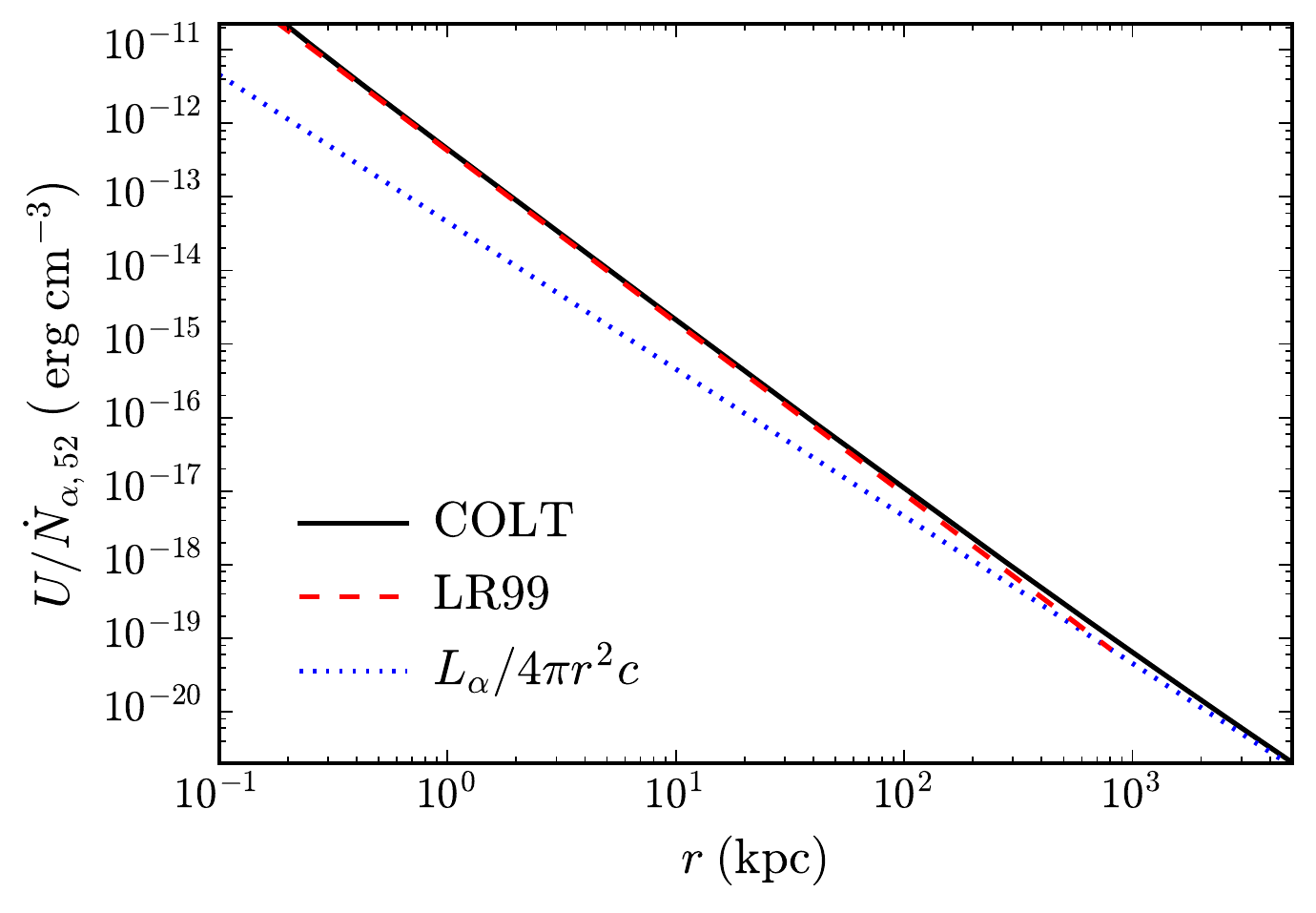}
    \caption{\protect\input{figures/LR99_Energy_Density/caption}}
    \label{fig:LR99_Energy_Density}
  \end{figure}

Figure~\ref{fig:LR99_Energy_Density} represents the energy density as a function of radius and demonstrates that the $\colt$ simulation smoothly connects the diffusion and free-streaming cases. The analytic expression for the angle-averaged intensity is
\begin{equation}
  \tilde{J} = \frac{1}{4 \upi} \left( \frac{9}{4 \upi \tilde{\nu}^3} \right)^{3/2} \exp \left\{ -\frac{9 \tilde{r}^2}{4 \tilde{\nu}^3} \right\} \, ,
\end{equation}
where the dimensionless radius, frequency, and intensity are given by $\tilde{r} \equiv r/r_\ast$,\ $\tilde{\nu} \equiv \nu/\nu_\ast$,\ and $\tilde{J} \equiv J/\left[ L_\alpha / (r_\ast^2 \nu_\ast) \right]$, respectively. The energy density in the Ly$\alpha$ radiation field is
\begin{align} \label{eq:LR99_Energy_Density}
  U(r) &= \frac{4 \upi}{c} \int_0^\infty J(\nu, r) d\nu \notag \\
       &= \frac{18^{1/3} \Gamma(\frac{13}{6})}{7 \upi^{3/2}} \frac{L_\alpha}{c\,r_\ast^2} \left( \frac{r_\ast}{r} \right)^{7/3} \notag \\
       &\approx 4.3 \times 10^{-13}~\text{erg~cm}^{-3}~\dot{N}_{\alpha, 52}~r_\text{kpc}^{-7/3} \, ,
\end{align}
which is independent of redshift. We have introduced the dimensionless quantity $\dot{N}_{\alpha, 52} \equiv \dot{N}_\alpha / (10^{52}$\,photons\,s$^{-1})$ to normalize the emission rate of the central Ly$\alpha$ source and the quantity $r_\text{kpc} \equiv r / (1\,\text{kpc})$ to normalize the radius. \hl{We note that the $r^{-7/3}$ scaling was also found by} \citet{Chuzhoy_2007}, \hl{who provide an intuitive explanation for the exponent.} This solution is plotted as the red dashed line and is valid out to $\sim 1$~Mpc where the redshifted photons transition to the optically thin limit. The free-streaming energy density is given by $L_\alpha / (4 \upi r^2 c)$ and is plotted as the blue dotted line in Fig.~\ref{fig:LR99_Energy_Density}. Scattering traps the photons and enhances their energy density at small radii compared to the optically thin case.

  \begin{figure}
    \centering
    \includegraphics[width=\columnwidth]{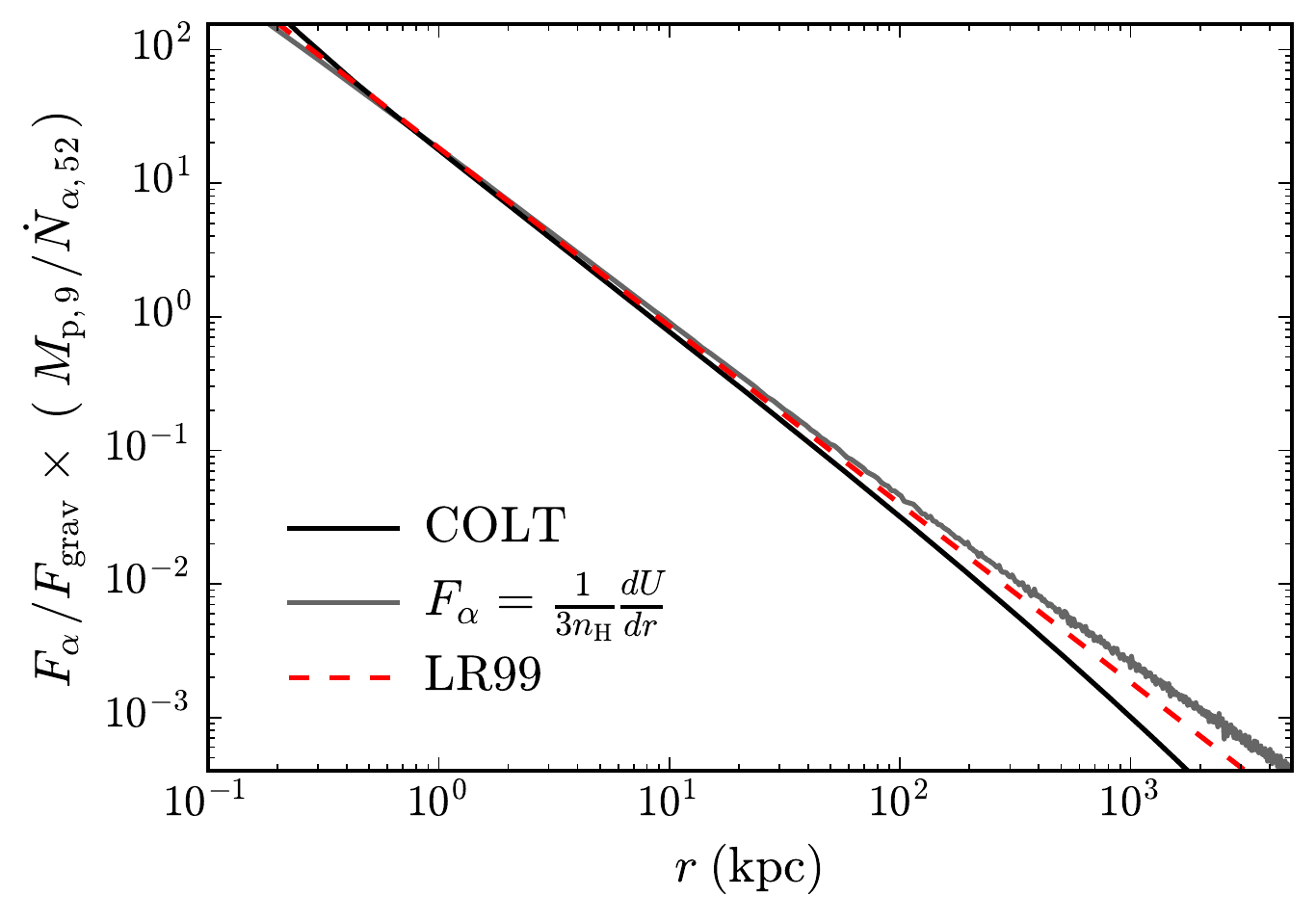}
    \caption{\protect\input{figures/LR99_Force_Ratio/caption}}
    \label{fig:LR99_Force_Ratio}
  \end{figure}

Following \citet{Dijkstra_Loeb_2008} we also compare the radiation force to the gravitational force as a function of radius in Fig.~\ref{fig:LR99_Force_Ratio}. For simplicity and consistency with the simulation setup we assume a point mass $M_\text{p}$ for the central source, which yields an inverse square law for the gravitational acceleration, $a_\text{grav} = G M_\text{p} / r^2$. The radiation force may be calculated directly by Monte-Carlo estimators as shown by the solid black curve. Additionally, in regions where the Eddington approximation holds we may calculate the acceleration due to Ly$\alpha$ feedback on the gas as
\begin{equation} \label{eq:LR99_a_rad}
  a_\alpha \approx \frac{1}{3 \rho_\text{H}} \frac{\text{d}U}{\text{d}r} \approx 80~\text{km~s}^{-1}\text{Myr}^{-1} \dot{N}_{\alpha, 52} r_\text{kpc}^{-10/3} \left( \frac{1+z}{11} \right)^{-3} .
\end{equation}
Thus, the ratio of the radiation force to the gravitational force is
\begin{align} \label{eq:LR99_Force_Ratio}
  \frac{F_\alpha}{F_\text{grav}}
    &\approx 18~\dot{N}_{\alpha, 52}~M_\text{p,9}^{-1}~r_\text{kpc}^{-4/3} \left( \frac{1+z}{11} \right)^{-3} \notag \\
    &\approx 4.3~\Upsilon_{\alpha,2}^{-1}~r_\text{kpc}^{-4/3} \left( \frac{1+z}{11} \right)^{-3} \, ,
\end{align}
where we have introduced $M_\text{p,9} \equiv M_\text{p} / (10^{9}~\Msun)$ as a normalization for the central point mass and $\Upsilon_\alpha \equiv (M_\text{p}/\Msun)/(L_\alpha/\Lsun)$ as the mass to (Ly$\alpha$) light ratio in solar units; furthermore, we define $\Upsilon_{\alpha,2} \equiv \Upsilon_\alpha/100$. Equation~(\ref{eq:LR99_Force_Ratio}) is plotted as the red dashed line in Fig.~\ref{fig:LR99_Force_Ratio}. The analytic and simulation curves in figure~2 of \citet{Dijkstra_Loeb_2008} bend downward at small radii compared to our case because the authors considered a modified NFW density profile for the dark matter halo. At large radii the Eddington approximation used to obtain the grey curve becomes increasingly unreliable. However, the different methods of estimating the Ly$\alpha$ radiation force are consistent with both the analytic expression and each other, in their region of respective validity.

\begin{figure}
    \centering
    \includegraphics[width=\columnwidth]{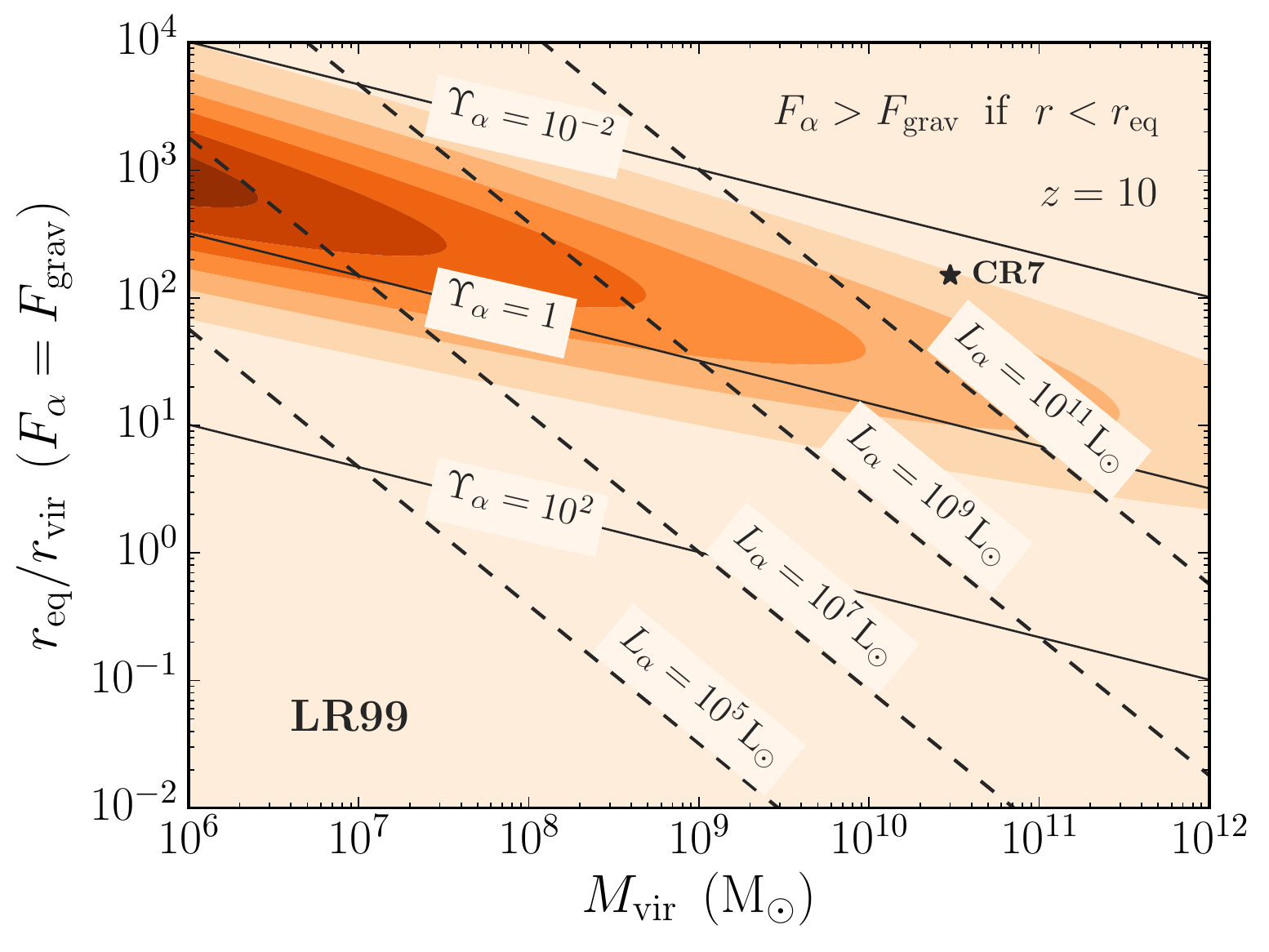}
    \caption{\protect\input{figures/LR99_Req_Rvir/caption}}
    \label{fig:LR99_Req_Rvir}
  \end{figure}

Finally, we investigate this model in the context of different galaxy scenarios. The Ly$\alpha$ radiation force overwhelms gravity within a characteristic radius corresponding to the equilibrium point where $F_\text{rad}/F_\text{grav} = 1$. In the case of Fig.~\ref{fig:LR99_Force_Ratio} the equilibrium radius is $r_\text{eq} \approx 9~\text{kpc}$ as calculated by the general expression $r_\text{eq} \approx 3~\text{kpc}\ \Upsilon_{\alpha,2}^{-3/4} [(1+z)/11]^{-9/4}$. The virial radius for a halo of mass $M_\text{vir}$ scales as $r_\text{vir} \propto M_\text{vir}^{1/3} (1+z)^{-1}$ so it follows that the ratio of the equilibrium radius to the virial radius is
\begin{equation} \label{eq:LR99_req_rvir}
  \frac{r_\text{eq}}{r_\text{vir}} \approx 14~\dot{N}_{\alpha, 52}^{3/4}~M_\text{p,9}^{-13/12} \left( \frac{1+z}{11} \right)^{-5/4} \, .
\end{equation}
Thus, assuming a constant mass to light ratio the potential for Ly$\alpha$-driven outflows is more substantial in less massive haloes. This is illustrated in Fig.~\ref{fig:LR99_Req_Rvir} for haloes at a redshift of $z = 10$. To better interpret lines of constant $\Upsilon_\alpha$ we include contours of probable values based on a universal star formation efficiency $f_\star$. If we assume the mass in baryons is $M_\text{b} \approx M_\text{vir} \Omega_\text{b} / \Omega_\text{m}$ and the mass in Pop~III stars is $M_\star \approx f_\star M_\text{b}$ then Equ.~(\ref{eq:LR99_req_rvir}) may be written as $r_\text{eq} / r_\text{vir} \approx 82\,[f_\star / (10^{-3})]^{3/4} M_\text{p,9}^{-1/3} [ (1+z)/11 ]^{-5/4}$. The contours represent the product of the root-mean-square amplitude of linearly extrapolated density fluctuations $\sigma$ \citep[see][]{Loeb_Furlanetto_2013} and a normal distribution on $\log f_\star$ with mean $\mu_{\log f_\star} = -3$ and standard deviation $\sigma_{\log f_\star} = 0.75$ in the vertical direction.

Although the Ly$\alpha$ force can have a greater relative impact than gravity it is important to realize the effect is predicated on the presence of neutral hydrogen gas. Thus, kinematic changes to galactic assembly may be mitigated if the ionization timescale~$t_\text{ion}$ is shorter than the minimum feedback timescale~$t_\alpha$. A simple order of magnitude estimate for $t_\alpha$ is the interval required to accelerate the gas to a velocity of $v_\alpha$, which according to Equation~(\ref{eq:LR99_a_rad}) is
\begin{equation}
  t_\alpha \sim \frac{v_\alpha}{a_\alpha} \approx 27~\text{kyr}~\frac{M_9^{10/9} v_{\alpha,10}}{\dot{N}_{\alpha, 54}} \left( \frac{r}{r_\text{vir}} \right)^{10/3} \left(\frac{1+z}{11}\right)^{-1/3} \, ,
\end{equation}
where the velocity normalization is $v_{\alpha,10} \equiv v_\alpha / (10~\text{km~s}^{-1})$, \hl{which is comparable to the thermal velocity of the gas.} In comparison the competing ionization timescale from the source is roughly $t_\text{ion} \sim N_e/\dot{N}_\text{ion}$. Here we approximate the number of available electrons in primordial gas by $N_e \approx (X + Y/2) M_b / m_H$, where $X \approx 0.75$ and $Y \approx 0.25$ represent the mass fraction of hydrogen and helium, respectively, and $M_b$ is the total mass of baryons being ionized, i.e. $M_b \approx M_\text{p} \Omega_b / \Omega_m$. The rate of ionizing photons is related to the rate of Ly$\alpha$ photons by $\dot{N}_\alpha \approx 0.68 (1 - f_\text{esc}) \dot{N}_\text{ion}$, where $f_\text{esc}$ is the escape fraction of ionizing photons \citep{Dijkstra_2014}. The above arguments combine to give
\begin{equation}
  t_\text{ion} \sim \frac{N_e}{\dot{N}_\text{ion}} \approx 327~\text{kyr}~\dot{N}_{\alpha, 52}^{-1}~M_\text{p,9} \,\left( 1 - \frac{f_\text{esc}}{0.08} \right) \, ,
\end{equation}
which demonstrates that it is possible for Ly$\alpha$ feedback to have a dynamical impact before the gas becomes ionized by the same central source. The Ly$\alpha$ timescale is about an order of magnitude shorter with strong radial dependence but exhibits only weak scaling with mass and redshift:
\begin{equation}
  \frac{t_\alpha}{t_\text{ion}} \approx 0.084~M_\text{p,9}^{1/9}~v_{\alpha,10} \,\left( \frac{r}{r_\text{vir}} \right)^{10/3} \left(\frac{1+z}{11}\right)^{-1/3} \, .
\end{equation}
The above calculation neglects recombinations which would slow the propagation of the ionization front. Also, the mass contained within a given radius could be significantly smaller than the halo mass. Still, as the estimates are based on a post-processing model, we defer further discussion until self-consistent dynamics are considered in Section~\ref{sec:dynamical_impact}.

\subsection{Overcoming the gravitational binding energy}
\label{sub:binding_energy}
While the previous discussion applies to neutral gas in the intergalactic medium we now consider a similar argument that connects to the virial halo itself. \citet{Rees_Ostriker_1977} considered a scenario in which Ly$\alpha$ cooling radiation overcomes the gravitational binding energy for a virialized system. The condition $L_\alpha t_\text{trap} \gtrsim G M^2 / R$ is roughly equivalent to the requirement that the trapping time be longer than the dynamic time, i.e. $t_\text{trap} > t_\text{dyn}$, where $t_\text{trap} \approx 15\,t_\text{light} \tau_6^{1/3} T_4^{1/6}$ for optical depths of $\tau_6 \equiv \tau / 10^6 \gtrsim 1$, and where $t_\text{light}$ is the light crossing time with $T_4 \equiv T / (10^4~\text{K})$ \citep{Adams_1975,Neufeld_1991}. This leads to a critical Ly$\alpha$ luminosity of \citep{Loeb_Furlanetto_2013}
\begin{equation}
  L_{\alpha, \text{crit}} \sim 10^{39}~\text{erg\,s}^{-1}~\left( \frac{M_\text{vir}}{10^6~\Msun} \right)^{4/3} \left( \frac{1+z}{11} \right)^2 \left( \frac{15\,t_\text{light}}{t_\text{trap}} \right) \, ,
\end{equation}
which is roughly equivalent to the luminosity generated by a $10^3~\Msun$ Population~III star cluster with an escape fraction for ionizing photons of $f_\text{esc} = 0.1$. \hl{A lower escape fraction implies that Ly$\alpha$ photons are more efficiently produced while a higher escape fraction reduces the Ly$\alpha$ luminosity for the same star formation efficiency~$f_\star$. The value of $f_\text{esc} = 0.1$ is a typical time-averaged escape fraction in $10^9~\Msun$ haloes, although $f_\text{esc}$ can be much higher for minihaloes ($\sim 0.5$) or much smaller for more mature haloes ($\lesssim$ a few per cent)} \citep[see e.g.][]{Gnedin_2008,Wise_2009,Yajima_2011,Ferrara_Loeb_2013,Paardekooper_2015,Faisst_2016,Xu_2016}. Thus, the maximum star formation efficiency to avoid unbinding via Ly$\alpha$ feedback is
\begin{equation}
  f_\star \equiv \frac{M_\star}{M_\text{gas}} \lesssim 10^{-3}~\left( \frac{M_\text{vir}}{10^6~\Msun} \right)^{1/3} \left( \frac{1+z}{11} \right)^2 \left( \frac{15\,t_\text{light}}{t_\text{trap}} \right) \, .
\end{equation}

\subsection{Ly$\alpha$-driven galactic supershells}
\label{sub:Lya_driven_galactic_supershells}
As previously mentioned, Ly$\alpha$ radiation pressure does not appreciably affect gas within ionized regions. We may estimate the kinematic effects of photon trapping in the context of the shell model by considering the following parameters: (i) the Ly$\alpha$ luminosity $L_\alpha$, (ii) the local \HI\ column density $N_\text{\HI}$, (iii) the shell radius $r$, and (iv) the duration that the gas remains neutral $\Delta t$. For simplicity we consider a geometry dependent force multiplier, $M_\text{F}$, as a way to absorb any uncertainty in the enhancement to the total force, $L_\alpha / c$, which is imparted by the source under single scattering. Therefore the change in velocity is approximately
\begin{align} \label{eq:v_alpha_shell}
  \Delta v &\approx \frac{M_\text{F} L_\alpha \Delta t}{4 \upi r^2 m_\text{H} N_\text{\HI} c} \notag \\
           &\approx 10~\text{km~s}^{-1}~M_\text{F,50}~L_{\alpha,8}~\Delta t_\text{kyr}~N_\text{\HI,19}^{-1}~r_{100\text{pc}}^{-2} \, ,
\end{align}
where we have used the following for notational convenience: $M_\text{F,50} \equiv M_\text{F} / 50$, $L_{\alpha,8} \equiv L_\alpha / (10^8~\Lsun)$, $\Delta t_\text{kyr} \equiv \Delta t / (1~\text{kyr})$, $N_\text{\HI,19} \equiv N_\text{\HI} / (10^{19}~\text{cm}^{-2})$, and $r_{100\text{pc}} \equiv r / (100~\text{pc})$. Even in the relatively short time of $\Delta t \approx 1~\text{kyr}$ the shell can accelerate to speeds comparable to the thermal velocity $v_\text{th} = 12.85~T_4^{1/2}~\text{km\;s}^{-1}$ and affect the local hydrodynamics.

\subsection{Impact of ionizing radiation pressure}
\label{sub:relative_impact_of_ionizing_radiation_pressure}
We now compare the impact of Ly$\alpha$ radiation pressure to that of ionizing radiation. To simplify the calculation we assume blackbody emission from a point source in the shell model considered in Section~\ref{sub:Lya_driven_galactic_supershells}. This spectrum may be parametrized by the bolometric luminosity~$L_\star$ and effective temperature~$T_\text{eff}$. Thus, the flux through a shell of radius~$r$ is given by \citep{Krumholz_Stone_2007,Greif_2009}
\begin{equation} \label{eq:Fnu_ion_blackbody}
  F_\nu = \frac{L_\star e^{-\tau_\nu}}{4 \sigma_\text{SB} T_\text{eff}^4 r^2} B_\nu \,
\end{equation}
where $\sigma_\text{SB}$ is the Stephan-Boltzmann constant, $B_\nu$ is the Planck function, and $\tau_\nu \approx \int_0^r \sigma_\nu n_\text{\HI} \text{d}\ell = \sigma_\nu N_\text{\HI}(r)$ is the optical depth to ionizing photons. If we also employ the nebular approximation then each photoionization occurs from the $1~^2S$ ground state of \HI\ and the frequency-dependent cross section becomes \citep[see equation 2.4 of][]{Osterbrock_book_2006}
\begin{equation} \label{eq:sigmanu_ion}
  \sigma_\nu = \sigma_0\,\left( \frac{13.6\,\text{eV}}{h\nu} \right)^4\,\frac{\exp[4 - 4_{\,} \varepsilon^{-1}\tan^{-1}\varepsilon]}{1 - \exp(-2 \pi / \varepsilon)} \, ,
\end{equation}
where $\varepsilon \equiv \sqrt{h \nu / 13.6~\text{eV} - 1}$ and $\sigma_0 \approx 6.3 \times 10^{-18}~\text{cm}^2$. For the gas coupling the absorption coefficient for ionizing radiation is $k_\nu = \rho \sigma_\nu / \mu_\text{H}$, where the mean molecular weight is roughly the mass of hydrogen, i.e. $\mu_\text{H} \approx m_\text{H}$. Putting this together we calculate the acceleration due to ionizing radiation
\begin{align} \label{eq:a_gamma_blackbody}
  a_\gamma &\equiv \frac{1}{c \rho} \int k_\nu F_\nu \text{d}\nu \notag \\
           &\approx \frac{L_\star}{4 c \mu_\text{H} \sigma_\text{SB} T_\text{eff}^4 r^2} \int_{\nu_\text{min}}^\infty \sigma_\nu B_\nu e^{-\tau_\nu} \text{d}\nu \notag \\
           &\approx \frac{15 \sigma_0 L_\star}{4 \pi^5 \mu_\text{H} r^2 c} \int_{x_\text{min}}^\infty \frac{x_\text{min}^4 \exp[4 - 4_{\,} \chi^{-1}\tan^{-1}\chi]}{x \left( e^x - 1 \right) \left[ 1 - \exp(-2 \pi / \chi) \right]} e^{-\tau_x} \text{d}x \notag \\
           &\approx \frac{15 \sigma_0 L_\star}{4 \pi^5 \mu_\text{H} r^2 c} \frac{x_\text{min}^3}{e^{x_\text{min}} - 1} \mathcal{T}_\gamma \notag \\
           &\approx 260~\text{km\,s}^{-1}\,\text{kyr}^{-1}~r_\text{pc}^{-2} \mathcal{T}_\gamma \, , 
\end{align}
where $\chi \equiv \sqrt{x / x_\text{min} - 1}$ and $x_\text{min} \equiv 13.6\,\text{eV}/(k_\text{B} T_\text{eff})$. \hl{The first line is the general definition while the second line incorporates the blackbody approximation from Equation}~\ref{eq:Fnu_ion_blackbody}. \hl{The third line employs a change of variables into nondimensional frequency and the cross section from Equation}~\ref{eq:sigmanu_ion} \hl{for direct integration. The fourth line takes advantage of the fact that at low $N_\text{\HI}(r)$ the integrand is sharply peaked at $x_\text{min}$ so we apply a delta function approximation to obtain the coefficient and introduce an ionizing radiation force transmission function $\mathcal{T}_\gamma \in [0,1]$ which we discuss later in this section.} The last line has been evaluated for a $100~\Msun$ Pop~III star with $L_\star = 10^{6.095}~\Lsun$ and $T_\text{eff} = 10^{4.975}~$K \citep{Schaerer_2002}. Finally, we may relate this to the Ly$\alpha$ radiation pressure by considering the ionizing photon rate \citep{Greif_2009}
\begin{equation}
  \dot{N}_\text{ion} = \frac{\pi L_\star}{\sigma_\text{SB} T_\text{eff}^4} \int_{\nu_\text{min}}^\infty \frac{B_\nu}{h \nu} \text{d}\nu = \frac{15 L_\star}{\pi^4 k_\text{B} T_\text{eff}} \int_{x_\text{min}}^\infty \frac{x^2 \text{d}x}{e^x - 1} \, .
\end{equation}
In connection to Equation~(\ref{eq:v_alpha_shell}), the acceleration $a_\alpha = \Delta v / \Delta t$ and Ly$\alpha$ photon rate $\dot{N}_\alpha \approx 0.68 (1 - f_\text{esc}) \dot{N}_\text{ion}$ provide the relative impact for a massive Pop~III star:
\begin{equation} \label{eq:a_alpha_gamma}
  \frac{a_\alpha}{a_\gamma} \approx 0.94~M_\text{F,50} N_\text{\HI,19}^{-1} \mathcal{T}_\gamma^{-1} \, .
\end{equation}
\hl{The Ly$\alpha$ force multiplier is certain to depend on the column density of the shell. However, the exact relationship may be complicated by several other factors such as geometry, bulk velocity, dust content, temperature, emission spectrum, and various three-dimensional effects such as turbulence and low-opacity holes. Analytic estimates suggest that $M_\text{F} \sim t_\text{trap}/t_\text{light} \approx 15\,(\tau_0/10^{5.5})^{1/3}$ for a uniform slab} \citep{Adams_1975}. \hl{In the context of Equations}~\ref{eq:v_alpha_shell}~\hl{and}~\ref{eq:a_alpha_gamma} \hl{we use the computed results from} \citet{Dijkstra_Loeb_2008} \hl{for static shells (see their figure~6). The data follow $M_\text{F} \approx 10\,N_\text{\HI,19}^{\,0.44}$ over the range $N_\text{\HI} \in (10^{19}, 10^{21})\,\text{cm}^{-2}$, indicating that the effective force ratio is bounded by $a_\alpha/a_\gamma \gtrsim 0.19\,N_\text{\HI,19}^{-0.56}$. Finally, we emphasize that this is the total column density of the shell while the final transmission term $\mathcal{T}_\gamma$ uses the cumulative radial column density (discussed below), therefore the ratio may change across the shell itself.}

In the idealized case of a thin, dense shell with no interior \HI\ it is possible that Ly$\alpha$ photons and ionizing radiation provide equal contributions to the acceleration. However, recombinations within the \HII\ region will absorb ionizing photons throughout the volume, yielding an effective \hl{ionizing radiation force transmission function that depends on the radial column density $N_\text{\HI}(r)$. The exact form of $\mathcal{T}_\gamma$ may be calculated according to Equation}~\ref{eq:a_gamma_blackbody}, \hl{however for intuition it may be thought of as a step function that rapidly decreases beyond a critical column density of $N_\text{\HI,crit} \approx 3 \times 10^{-17}\,\text{cm}^{-2}$. If the vanishing transmission is modeled as an effective optical depth we find $\mathcal{T}_\gamma \approx \exp[-2.3 \times 10^{-18}\,\text{cm}^2\,N_\text{\HI}]$ but this falls off much faster than the exact solution, so another reasonable approximation for $N_\text{\HI} \in (10^{17}, 10^{21})$~cm$^{-2}$ is the complementary error function $\mathcal{T}_\gamma \approx \text{erfc}[1.15 \log(N_\text{\HI}/N_\text{\HI,crit})]/2$.} As the ionization front advances this recombination opacity increases until it balances the ionization rate. Thus, as long as the shell remains neutral we expect the acceleration to be dominated by Ly$\alpha$ radiation pressure. For concreteness, under Case~B recombination we may estimate the optical depth as $\tau_\nu \sim (\alpha_\text{B} n^2 \text{d}t) \sigma_0 \text{d}\ell \sim 10\,\text{d}\ell_\text{S}^2 (\dot{N}_\text{ion,50} n_{10})^{2/3}$. Here, $\dot{N}_\text{ion,50} \equiv \dot{N}_\text{ion} / (10^{50}s^{-1})$ and $n_{10} \equiv n / (10\,\text{cm}^{-3})$, where we have used the light crossing time $\text{d}t \sim \text{d}\ell/c$. Furthermore, the distance is scaled to the Str{\"o}mgren radius, i.e. $\text{d}\ell_\text{S} \equiv \text{d}\ell / R_\text{S}$. With these values Ly$\alpha$ momentum transfer exceeds that of ionizing radiation at the ionization front by several orders of magnitude.

\section{Methodology: Radiative Transfer}
\label{sec:radiative_transfer_methodology}
In this section we discuss the post-processing methodology that enables us to accurately calculate the Ly$\alpha$ radiation pressure. For details regarding the MCRT method employed in the Cosmic Ly$\alpha$ Transfer code ($\colt$) the reader is referred to \citet{Smith_2015}. The numerical methods outlined below introduce additional functionality for $\colt$ that is then tested in preparation for fully coupled radiation hydrodynamics. Although $\colt$ is capable of calculating radiation pressure for general three-dimensional grids with mesh refinement we focus on spherically symmetric profiles for this paper. In Sections~\ref{sub:energy_density_due_to_photon_trapping}, \ref{sub:force_due_to_scattering_events}, and \ref{sub:pressure_due_to_momentum_flux} we describe methods for calculating the Ly$\alpha$ energy density, force, and radiation pressure, respectively.

\subsection{Energy density due to photon trapping}
\label{sub:energy_density_due_to_photon_trapping}
The Monte-Carlo method naturally tracks the time photons spend within each cell, which may be used as a measure of the energy density in the Ly$\alpha$ radiation field. This is because the total number of photons in a cell at a given time is reasonably approximated by the Ly$\alpha$ emission rate multiplied by the average time spent in the cell, i.e. $N_\text{cell} = \dot{N}_\alpha \langle t \rangle_\text{cell}$, which is related to the Ly$\alpha$ source luminosity via $L_\alpha = h \nu \dot{N}_\alpha$. For convenience we assume all photons are relatively close in energy, i.e. $\Delta \nu \ll \nu_\alpha$,\footnote{For line transfer this is justified because even photons in the extreme wing of the profile differ from the median energy by a small fraction, e.g. a velocity offset of $\Delta v \approx 3000~\text{km~s}^{-1}$ corresponds to a one per cent deviation from the assumed energy $h \nu_\alpha$.} which is straightforward to relax if desired. In this framework we keep track of the total length traversed by all paths of all photons within the particular cell and define $\sum \ell_\text{cell} \equiv c N_\text{ph} \langle t \rangle_\text{cell}$ where $N_\text{ph}$ is the number of photon packets in the simulation. The energy density is obtained after integrating over all independent photon paths up to the point of escape:
\begin{equation} \label{eq:U_cell}
  U_\text{cell} = \frac{h\nu N_\text{cell}}{V_\text{cell}}
                = \frac{L_\alpha}{c N_\text{ph} V_\text{cell}} \sum \ell_\text{cell} \, .
\end{equation}
The summation is over all paths of all photons within the particular cell and the volume depends on the geometry under consideration,
e.g. $V_\text{cell} = \Delta x \Delta y \Delta z$ for Cartesian cells and $V_\text{cell} = \frac{4 \upi}{3} (r^3_\text{outer} - r^3_\text{inner})$ for spherical shells. Finally, the sum over the average residence time is equivalent to the total trapping time over the cells in a domain $\mathcal D$, i.e. $t_\text{trap} = N_\text{ph} \sum \langle t \rangle_{\mathcal D}$.

\subsection{Radiation force due to scattering events}
\label{sub:force_due_to_scattering_events}
Monte-Carlo radiative transfer also lends itself to tracking momentum transfer from Ly$\alpha$ photon packets to hydrogen atoms during each scattering event. The exchange of momentum is given by $\Delta\bmath{p} = \frac{h\nu}{c} ( \bmath{n}_i - \bmath{n}_f )$, where the unit vectors $\bmath{n}_i$ and $\bmath{n}_f$ represent the initial and final directions of the scattered photon. For convenience we track the overall momentum exchange which is related to the average by $\sum \Delta \bmath{p} = N_\text{ph} \langle \Delta \bmath{p} \rangle$. To good approximation the momentum rate $\dot{\bmath{p}} \equiv \Delta \bmath{p} / \Delta t$ may be written in terms of the Ly$\alpha$ emission rate~$\dot{N}_\alpha$. However, the effect of each scattering event must be applied to the entire cell as a single fluid element. Therefore, the force is diluted by the inertial mass of the cell, i.e. $m_\text{cell} = \int \rho dV$. The acceleration due to Ly$\alpha$ scattering is
\begin{equation} \label{eq:F_cell}
  \bmath{a}_\text{cell} = \frac{\dot{N}_\alpha \langle \Delta \bmath{p} \rangle}{\rho V_\text{cell}}
                        = \frac{L_\alpha}{c N_\text{ph} \rho V_\text{cell}}
                          \sum \left( \bmath{n}_i - \bmath{n}_f \right) \, ,
\end{equation}
where the summation is over all scatterings of all photons within the particular cell. The geometry determines the direction of interest for the contribution from each scattering event. For example, the radial component of momentum is $\Delta p_r \propto (\bmath{n}_i - \bmath{n}_f) \bmath{\cdot} \hat{\bmath{r}}$. In practice we employ a Monte-Carlo estimator to reduce noise in regions with fewer scatterings. The idea is to use continuous momentum deposition to determine the gas coupling. The photon contribution is weighted by the traversed optical depth so the acceleration becomes
\begin{equation} \label{eq:F_cell_MCE}
  \bmath{a}_\text{cell} = \frac{L_\alpha}{c N_\text{ph} \rho V_\text{cell}}
                          \sum d\tau_\nu \bmath{n} \, ,
\end{equation}
where the sum is over all photon paths within the cell.

\subsection{Radiation pressure due to the rate of momentum flux}
\label{sub:pressure_due_to_momentum_flux}
The final quantity of interest is Ly$\alpha$ radiation pressure, which differs from the energy density and radiation force as it applies to cell boundaries rather than interiors. Therefore, we consider the rate of momentum flux as measured by the stream of photons propagating through a given cell surface. Similar to the force calculation the momentum of the photon packet is $\bmath{p} = \frac{h\nu}{c} \bmath{n}$, which we can imagine is transferred to a theoretical barrier representing the surface. Therefore, the Ly$\alpha$ radiation pressure tensor is
\begin{equation} \label{eq:P_cell}
  \textbfss{P}_\text{cell} = \frac{\dot{N}_\alpha \langle \bmath{p} \otimes \bmath{n} \rangle}{A_\text{cell}}
      = \frac{L_\alpha}{c N_\text{ph} A_\text{cell}}
        \sum \bmath{n} \otimes \bmath{n} \, ,
\end{equation}
where the tensor product of the direction vector with itself is a symmetric matrix, i.e. $\bmath{n} \otimes \bmath{n} = \bmath{n} \bmath{n}^\text{T}$. Here $A_\text{cell}$ is the area of the cell boundary, e.g. a Cartesian cell has $A_x = \Delta y \Delta z$ and a spherical shell has $A_r = 4 \upi r^2$. The summation is over all crossings of all photons through the particular interface. In practice we again employ a Monte-Carlo estimator to calculate the continuous momentum flux in a particular direction. The pressure tensor becomes
\begin{equation} \label{eq:P_cell_MCE}
  \textbfss{P}_\text{cell} = \frac{L_\alpha}{c N_\text{ph} V_\text{cell}} \sum \ell_\text{cell} d\tau_\nu \bmath{n} \otimes \bmath{n} \, ,
\end{equation}
where the sum is over all photon paths within the cell and $\ell_\text{cell}$ is the distance traversed. For concreteness, to determine the radial pressure in spherical shells the projected outer product terms reduce to $\Delta \textbfss{P}_{rr} \propto (\bmath{n} \otimes \bmath{n})_{rr} = \bmath{n}_r^2 = (\bmath{n} \bmath{\cdot} \hat{\bmath{r}})^2$.

\subsection{Test Cases}
\label{sec:tests}
Before proceeding further we verify $\colt$ against known solutions. This was done in Section~\ref{sub:sources_in_a_neutral_comoving_igm} for the case of sources in an expanding IGM. For completeness we also compare the Monte-Carlo estimators for $U$ and $\textbfss{P}$ in the diffusion and free streaming limits.

\subsubsection{A point source in vacuum}
\label{sub:a_point_source_in_vacuum}
A point source in vacuum is not expected to undergo any scattering events. Thus, a free streaming situation arises where the energy in a shell at radius $r = ct$ remains constant in time but is spread through an ever increasing volume of $\Delta V = 4 \upi r^2 \Delta r$, where $\Delta r$ is the shell thickness. The shell flash has an energy of $\Delta E = L_\alpha \Delta t = L_\alpha \Delta r / c$ and thus an energy density of $U(r) = L_\alpha / (4 \upi r^2 c)$, as verified by $\colt$. Also, with no scattering events the force is zero and the relation to pressure is $P = U$ as expected in this special case.

\subsubsection{Energy density in the diffusion limit}
\label{sub:energy_density_in_the_diffusion_limit}
\begin{figure}
    \centering
    \includegraphics[width=\columnwidth]{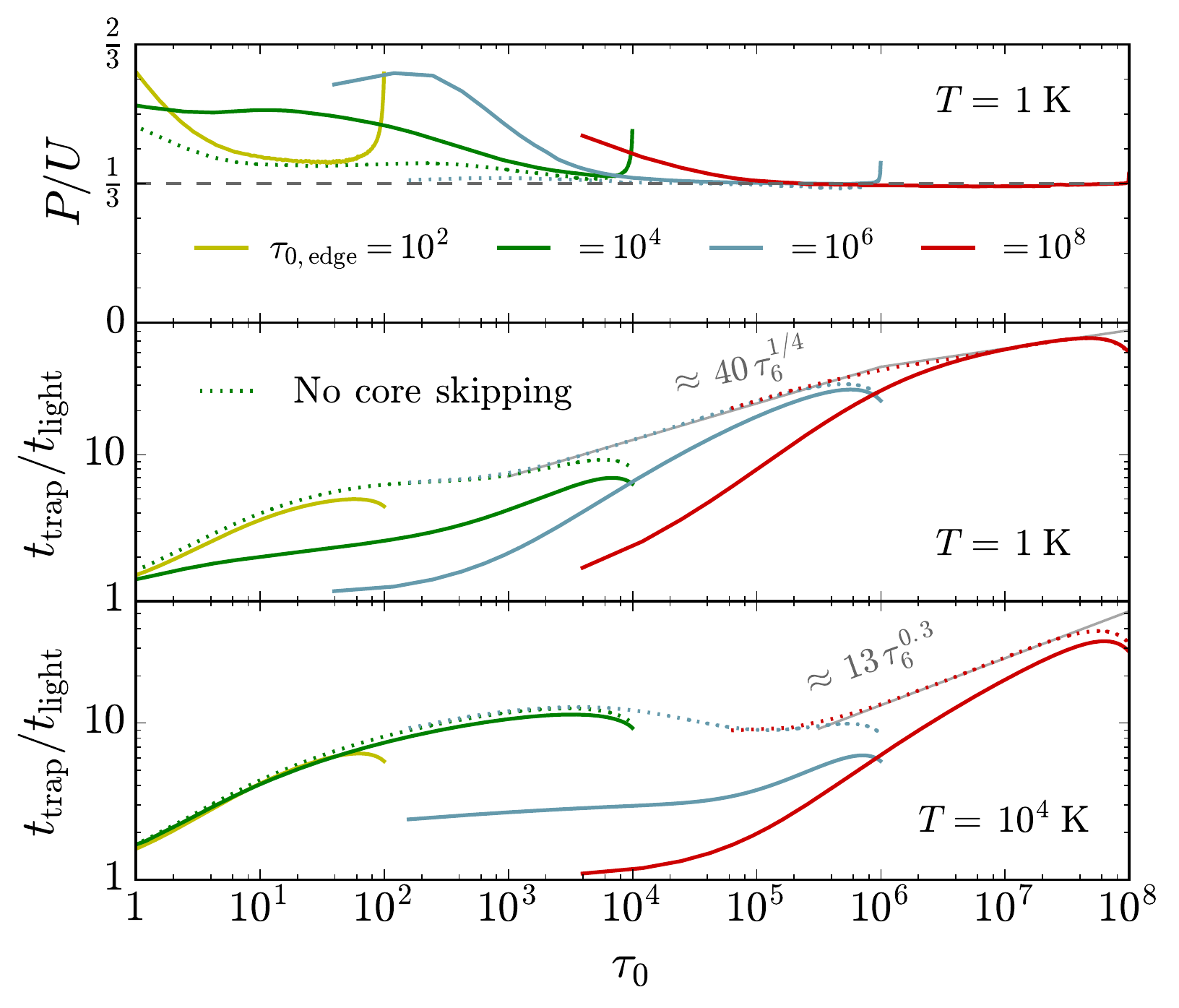}
    \caption{\protect\input{figures/diffusion_test/caption}}
    \label{fig:diffusion_test}
  \end{figure}

Photons in optically thick environments are expected to undergo numerous scattering events. In these conditions the isotropic pressure and energy density are related as $P = U / 3$. We perform this test on uniform density spheres with various centre-to-edge optical depths $\tau_{0,\text{edge}} = \{10^2, 10^4, 10^6, 10^8\}$ at low \hl{and high} temperature\hl{s}, i.e. $T = 1$~K \hl{and $T = 10^4$~K}. The pressure to energy density ratio and trapping time normalized to the light crossing time are shown in Fig.~\ref{fig:diffusion_test}. The solid (dotted) lines are simulations with (without) the core skipping scheme, which significantly improves the computational efficiency but underestimates the trapping time for the innermost radial shells. The trapping time is approximately $t_\text{trap,1\,K} \approx \{1.5, 3.9, 6.3, 7.5, 13, 23, 35, 52, 70\}$ \hl{and $t_{\text{trap},10^4\,\text{K}} \approx \{1.6, 4.2, 8.2, 12, 12, 9.1, 12, 25, 50\}$} light crossings at the radial optical depths $\log(\tau_0) = \{0, 1, 2, 3, 4, 5, 6, 7, 8\}$ \hl{for $T = 1$~K and $T = 10^4$~K, respectively. At $T = 1$~K} this is approximately fit by $t_\text{trap} / t_\text{light} \approx 40 (\tau_0 / 10^6)^{1/4}$ over the interval $10^3 < \tau_0 < 10^6$ although the slope becomes more shallow ($\approx 1/8$) for $\tau_0 > 10^6$. \hl{At $T = 10^4$~K the fit is roughly given by $t_\text{trap} / t_\text{light} \approx 13 (\tau_0 / 10^6)^{0.3}$ for $\tau_0 \gtrsim 10^6$, although we note it is robustly $\gtrsim 10$ for $\tau_0 \gtrsim 100$.}

\hl{It is particularly interesting that the ratio of $t_\text{trap} / t_\text{light}$ scales more slowly than predicted by analytic arguments} \citep{Adams_1975}. \hl{The standard $\tau_0^{1/3}$ scaling assumes that Ly$\alpha$ escapes in a `single excursion' during which there is a `restoring force' that pushes Ly$\alpha$ wing photons back to the core by an amount $\langle \text{d}x | x \rangle \approx -1 / |x|$ for each scattering event, where $x \equiv (\nu - \nu_\alpha)/\Delta \nu_\text{D}$ is the relative frequency from line centre in units of Doppler widths $\Delta \nu_\text{D} \equiv (v_\text{th}/c) \nu_\alpha \approx 10^{11}~\text{Hz}~T_4^{1/2}$} \citep[see][]{Adams_1972,Harrington_1973}. \hl{However, in cold media and far in the wing of the line, this restoring force becomes smaller than the `force' exerted by atomic recoil. Specifically, recoil pushes Lya photons consistently to the red by an amount $\text{d}x \approx g \equiv h \Delta \nu_\text{D} / 2 k_\text{B} T \approx 0.02536/\sqrt{T/\text{K}}$, where $g$ is known as the `recoil parameter'} \citep[see e.g.][]{Field_1959,Adams_1971}. \hl{For $x \gtrsim g^{-1} \approx 39.44 \sqrt{T/\text{K}}$ we expect recoil to overwhelm the standard `restoring force'. In comparison, the typical frequency with which photons escape from a medium with a given optical depth and temperature is $x_\text{peak} \approx 33.6\,(\tau_0 / 10^6)^{1/3}(T/\text{K})^{-1/6}$, see the discussion following equation 34 in} \citet{Smith_2015} \hl{based on the analytic solution for a static sphere derived by} \citet{Dijkstra_2006}. \hl{In short, for $\tau_0 \gtrsim 1.62 \times 10^6 (T/\text{K})^2$ photons need to diffuse so far into the wing of the line in order to escape that recoil prevents them from returning to the core, which further facilitates their escape and likely explains the flattening of the trapping time at higher $\tau_0$. Interestingly, at $T = 10^4$~K we find the trapping time is closer to the analytic estimate, although the scaling exponent is still only $\sim 0.3$.} We also note that this result depends on \hl{other} environmental properties such as geometry.

\section{Methodology: Radiation Hydrodynamics }
\label{sec:radiation_hydrodynamics_methodology}
Although there are several descriptions detailing the theory of radiation hydrodynamics, our treatment closely follows the conventions found in \citet{Mihalas_book_1984}, \citet{Mihalas_2001}, and \citet{Castor_book_2004}. We also benefited from the work of \citet{Abdikamalov_2012}, \citet{Harries_2015}, \citet{Roth_2015}, and \citet{Tsang_2015}.

\subsection{Ly$\alpha$ radiative transport}
The lab frame Ly$\alpha$ radiative transfer equation is
\begin{equation} \label{eq:transfer_equation}
  \frac{1}{c} \frac{\upartial I_\nu}{\upartial t} + \bmath{n} \bmath{\cdot} \bmath{\nabla} I_\nu = k_\nu \left( J_\nu - I_\nu \right) + S_\nu(\bmath{r}) \, ,
\end{equation}
where $I_\nu$ is the specific intensity of the radiation, $\nu$ is the frequency, $\bmath{n}$ is a unit vector in an arbitrary direction, and $J_\nu \equiv \frac{1}{4 \upi} \int I_\nu \text{d}\Omega$ is the intensity averaged over the solid angle $\text{d}\Omega$. The source terms on the right are represented by $S_\nu(\bmath{r})$, the emission function for newly created photons at the position $\bmath{r}$, and $k_\nu \equiv n_\text{\HI} \sigma_\nu$, the absorption coefficient which is equivalent to the neutral hydrogen number density multiplied by the frequency-dependent Ly$\alpha$ scattering cross-section. Additional terms may be included to describe absorption and scattering of Ly$\alpha$ photons by dust as well as other thermal processes. However, in this paper we consider only `dust-free' environments so we do not include a detailed description here.

The moments of the radiation intensity correspond to the Ly$\alpha$ energy density $E_\nu \equiv c^{-1} \int \text{d}\Omega I_\nu$, flux $\bmath{F}_\nu \equiv \int \text{d}\Omega I_\nu \bmath{n}$, and pressure $\textbfss{P}_\nu \equiv c^{-1} \int \text{d}\Omega I_\nu \bmath{n} \otimes \bmath{n}$. Each of these quantities may be integrated over frequency to obtain bolometric versions, e.g. $E = \int_0^\infty E_\nu \text{d}\nu$. With the intensity moments in hand we may now take moments of Equation~(\ref{eq:transfer_equation}) to arrive at the radiation energy and radiation momentum equations:
\begin{align}
  \label{eq:transfer_moments_1}
  &\frac{\upartial E}{\upartial t} + \bmath{\nabla} \bmath{\cdot} \bmath{F} = \iint \text{d}\nu \text{d}\Omega \, k_\nu \left( J_\nu - I_\nu \right) \equiv -c G^0 \, , \\
  \label{eq:transfer_moments_2}
  &\frac{1}{c^2} \frac{\upartial \bmath{F}}{\upartial t} + \bmath{\nabla} \bmath{\cdot} \textbfss{P} = \frac{1}{c} \iint \text{d}\nu \text{d}\Omega \, k_\nu \left( J_\nu - I_\nu \right) \bmath{n} \equiv -\bmath{G} \, .
\end{align}
These represent sources of energy and momentum coupling to the gas and thus are identified by $-c G^0$ and $-\bmath{G}$, respectively. In Equations~(\ref{eq:transfer_moments_1})~and~(\ref{eq:transfer_moments_2}) we have dropped the term corresponding to the Ly$\alpha$ emission function $S_\nu(\bmath{r})$, which in general should be included. This was done because throughout this paper we assume central emission which may be approximated as $S_\nu(\bmath{r}) \approx h \nu_\alpha \dot{N}_\alpha (4 \upi)^{-1} \delta(\nu) \delta(\bmath{r})$ and implies an energy source of $\iint \text{d}\nu \text{d}\Omega \, S_\nu(\bmath{r}) \approx L_\alpha \delta(\bmath{r})$. However, these photons originate from the ionizing source directly so there is no corresponding term for the gas in our simplistic modeling of an unresolved central Ly$\alpha$ source. We note that the non-equilibrium chemistry and cooling discussed in Section~\ref{sec:non-equilibrium_chemistry} does include radiative processes involving the gas at grid scales, such as collisional (de)excitation.

\subsection{Lab frame radiation hydrodynamics}
The equations governing non-relativistic hydrodynamics are often written in an Eularian reference frame as a set of conservation laws. We quote the conservation of mass, momentum, and total energy with 
source terms related to gravity and radiation as follows:
\begin{align}
  \label{eq:Eularian_RHD_1}
  &\frac{\upartial \rho}{\upartial t} + \bmath{\nabla} \bmath{\cdot} \left(\rho \bmath{v} \right) = 0 \, , \\
  \label{eq:Eularian_RHD_2}
  &\frac{\upartial \rho \bmath{v}}{\upartial t} + \bmath{\nabla} \bmath{\cdot} \left( \rho \bmath{v} \otimes \bmath{v} \right) + \bmath{\nabla} P_0 = -\rho \bmath{\nabla} \Phi + \bmath{G} - \frac{\bmath{v}}{c} G^0 \, , \\
  \label{eq:Eularian_RHD_3}
  &\frac{\upartial \rho e}{\upartial t} + \bmath{\nabla} \bmath{\cdot} \left[ \left( \rho e + P_0 \right) \bmath{v} \right] = -\rho \bmath{v} \bmath{\cdot} \bmath{\nabla} \Phi + c G^0 \, .
\end{align}
Here and in the following, all quantities with the subscript $0$ are evaluated in the comoving fluid reference frame while all other quantities are in the lab frame, or the Eularian frame of the fixed coordinate system. Therefore, $\rho$ is the lab frame density, $\bmath{v}$ is the lab frame velocity, $P_0$ is the comoving pressure, and the last equation is written in terms of the total specific energy, or the sum of comoving specific internal energy and lab frame specific kinetic energy: $e \equiv \epsilon_0 + \frac{1}{2} |\bmath{v}|^2$. The radiation source terms $\bmath{G}$ and $c G^0$ are respectively the momentum and energy components of the lab frame force four-vector, which directly correspond to the source terms of Equations~(\ref{eq:transfer_moments_1})~and~(\ref{eq:transfer_moments_2}).

In this paper we assume an ideal gas equation of state so that the comoving pressure is isotropic and completely specified by
\begin{equation}
  P_0 = (\gamma_\text{ad} - 1) \rho \epsilon_0 \, ,
\end{equation}
where $\gamma_\text{ad} \equiv C_P/C_V$ is the adiabatic index, or ratio of specific heat at constant pressure to that at constant volume.

The final source term we have included is gravity, which is represented by the scalar potential $\Phi$ and may be a function of space and time. However, in spherical symmetry we may utilize Newton's shell theorem. Thus, the acceleration depends on the mass enclosed within the radius of the shell, i.e. $\bmath{a}_\text{grav} = -\bmath{\nabla} \Phi = -G M_{<r} \hat{\bmath{r}}/r^2$. In the context of a flat $\Lambda$CDM cosmology there is also a background potential specified by $\nabla^2 \Phi = 4 \upi G \rho - \Lambda$ where the cosmological constant is given by $\Lambda \equiv 8 \upi G \rho_\Lambda = 3 H_0^2 \Omega_\Lambda$. Therefore, in cosmological environments we include deceleration from dark matter and cosmic acceleration from dark energy. In the simplest scenario this corresponds to adding the non-baryonic matter component to the enclosed mass $M_{<r}$ and an outward acceleration of $\bmath{a}_\Lambda = - \bmath{\nabla} \Phi_\Lambda = \frac{\Lambda}{3} \bmath{r}$. This allows us to employ physical units for the remainder of this paper even when cosmological dynamics are important.

\subsection{Ionizing radiation}
A proper treatment of ionizing radiation is essential for assessing the dynamical impact of Ly$\alpha$ feedback. We therefore solve the time-dependent transfer equation for ionizing radiation in an explicitly photon-conserving manner \citep{Abel_1999,Pawlik_2011}. This approach guarantees that the ionization fronts move at the correct speed independent of the spatial resolution. The ionizing radiation is characterized by expectation values of photons collected into three bands according to the ionization energies of \HI, \HeI, and \HeII, or 13.6~eV, 24.6~eV, and 54.4~eV, respectively. The absorption within each band is treated in the grey approximation using the source intensity $J_\nu$ and photoionization cross-sections $\sigma_\text{\HI}$, $\sigma_\text{\HeI}$, and $\sigma_\text{\HeII}$. The rate of ionizing photons in each frequency range is \citep{Jeon_XRB_2014}
\begin{equation} \label{eq:Ndot_ion}
  \dot{N}_{\text{ion},i} \equiv \int_{\nu_{\text{min},i}}^{\nu_{\text{max},i}} d\nu \frac{4 \pi J_\nu}{h \nu} \, ,
\end{equation}
where the subscript $i \in$~\{1, 2, 3\} denotes the particular band and $\dot{N}_\text{ion} = \sum \dot{N}_{\text{ion},i}$. The photoionization rate is defined by
\begin{equation}
  \Gamma_{x,i} \equiv \int_{\nu_{\text{min},i}}^{\nu_{\text{max},i}} d\nu \frac{4 \pi J_\nu}{h \nu} \sigma_x \, ,
\end{equation}
while the photoheating rate is given by
\begin{equation}
  \mathcal{E}_{x,i} \equiv \int_{\nu_{\text{min},i}}^{\nu_{\text{max},i}} d\nu \frac{4 \pi J_\nu}{h \nu} \sigma_x (h \nu - h \nu_x) \, ,
\end{equation}
where the subscript $x \in$~\{\HI, \HeI, \HeII\} denotes the species. Therefore, the average photoionization cross-section and average energy imparted to the gas per ionizing photon is
\begin{equation} \label{eq:sigma_ion}
  \langle \sigma_x \rangle_i \equiv \frac{\Gamma_{x,i}}{\dot{N}_{\text{ion},i}} \qquad \text{and} \qquad
  \langle \varepsilon_x \rangle_i \equiv \frac{\mathcal{E}_{x,i}}{\Gamma_{x,i}} \, .
\end{equation}

To guarantee conservation of photons over a hydrodynamical timestep $\Delta t$ we calculate the total number of ionizing photons emitted by the source as $N_{\text{ion},i} = \dot{N}_{\text{ion},i} \Delta t$. However, as photons are absorbed near the source the effective number of ionizing photons $N_{\text{ion,eff},i}$ is reduced to represent the remaining portion. Therefore, due to absorption in interior cells the effective photon rate is $\dot{N}_{\text{ion,eff},i} = N_{\text{ion,eff},i} / \Delta t$. For a shell of thickness $\Delta r$ the optical depth for a given species and frequency band is $\tau_{x,i} = n_x \langle \sigma_x \rangle_i \Delta r$ while the total optical depth from all species is $\tau_i = \sum \tau_{x,i}$. Therefore, the absorption rate in the cell is
\begin{equation}
  \dot{N}_{\text{ion,abs},i} = \dot{N}_{\text{ion,eff},i} \left( 1 - e^{-\tau_i} \right) \, .
\end{equation}
The rate of change in number density is
\begin{equation}
  \dot{n}_{\text{ion},x} = \sum_i \dot{n}_{\text{ion},x,i} = \sum_i \frac{\tau_{x,i}}{\tau_i} \frac{\dot{N}_{\text{ion,abs},i}}{V_\text{cell}} \, ,
\end{equation}
the rate of change in specific internal energy by volume is
\begin{equation}
  \dot{\Gamma} = \sum_{x,i} \langle \varepsilon_x \rangle_i \dot{n}_{\text{ion},x,i} \, ,
\end{equation}
and the acceleration along the ray due to ionizing momentum transfer is
\begin{equation} \label{eq:a_gamma}
  a_\gamma = \sum_{x,i} \frac{\langle h \nu_x \rangle_i}{c} \frac{\dot{n}_{\text{ion},x,i}}{\rho} \, .
\end{equation}
where $\langle h \nu_x \rangle_i \equiv h \nu_x + \langle \varepsilon_x \rangle_i$ is the average photon energy.

\subsection{Non-equilibrium chemistry and cooling} \label{sec:non-equilibrium_chemistry}
We self-consistently solve the rate equations for a primordial chemistry network consisting of H, H${}^+$, He, He${}^+$, He${}^{++}$, and e$^{-}$ \citep{Bromm_2002}. Reactions affecting these abundances include H and He collisional ionization and recombination from \citet{Cen_1992}. After incorporating the photoionization rates we obtain overall number density rates, i.e. $\dot{n}_x = \dot{n}_{\text{ion},x} + \dot{n}_{\text{chem},x}$. Meanwhile the cooling mechanisms include collisional ionization, collisional excitation, recombination cooling, bremsstrahlung, and inverse Compton cooling. Collectively, the cooling provides the rate of change in specific internal energy by volume $\dot{\Lambda}$. For computational efficiency these rates are tabulated as a function of the logarithmic temperature and linear interpolation is employed throughout the simulation.

\subsection{Timestep criteria} \label{sec:timestep_criteria}
Because photoionization can rapidly affect the chemical and thermal state of the gas we employ timestep sub-cycling $\Delta t_\text{sub}$ to accurately follow the evolution. We employ an explicit algorithm with timesteps limited to a small fraction of the cooling timescale and depletion timescales for individual abundances:
\begin{equation}
  \Delta t_\text{sub} = \epsilon_\text{sub} \min_x \left\{ \bigg|\frac{\epsilon_0 \rho}{\dot{\Gamma} - \dot{\Lambda}}\bigg|, \bigg|\frac{n_x}{\dot{n}_x}\bigg| \right\} \, ,
\end{equation}
where $\epsilon_\text{sub} \lesssim 0.1$ is the sub-cycling factor. It is also important to ensure the sub-cycling stops after the hydrodynamical time $\Delta t$. Also the number of absorptions cannot exceed the number of available ionizing photons, i.e. $\Delta t_\text{sub} \leq N_{\text{ion,eff},i} / \dot{N}_{\text{ion,abs},i}$. Therefore, the (volume) specific heating and cooling in each sub-cycle are $\Gamma = \dot{\Gamma} \Delta t_\text{sub}$ and $\Lambda = \dot{\Lambda} \Delta t_\text{sub}$, respectively. To ensure the accuracy of the chemical and thermodynamic state of the gas we update the species abundances and internal energy throughout each sub-cycle. However, we apply the cumulative force from the absorption of ionizing photons at the end of the hydrodynamical timestep. Therefore, the acceleration due to ionizing radiation is the weighted average from each sub-timestep, i.e. $a_\gamma = \sum a_{\gamma,\text{sub}} \Delta t_\text{sub} / \Delta t$.

\subsection{Monte-Carlo estimators} \label{sec:Monte-Carlo_estimators}
The final ingredient is to evaluate $G^0$ and $\bmath{G}$ in terms of the Monte-Carlo estimators discussed in Section~\ref{sec:radiative_transfer_methodology}. The MCRT calculations take place in the comoving frame and are related to quantities in the lab frame by the Lorentz transformation \citep{Mihalas_2001}:
\begin{align}
  G^0 &= \gamma \left( G_0^0 + \frac{\bmath{v}}{c} \bmath{\cdot} \bmath{G}_0 \right) \approx G_0^0 + \frac{\bmath{v}}{c} \bmath{\cdot} \bmath{G}_0 \, , \\
  \bmath{G} &= \bmath{G}_0 + \gamma \frac{\bmath{v}}{c} \left( G_0^0 + \frac{\gamma}{\gamma + 1} \frac{\bmath{v}}{c} \bmath{\cdot} \bmath{G}_0 \right) \approx \bmath{G}_0 + \frac{\bmath{v}}{c} \bmath{\cdot} \bmath{G}_0 \, ,
\end{align}
where $\gamma \equiv (1 - |\bmath{v}|^2 / c^2)^{-1/2}$. Although Ly$\alpha$ scattering is only approximately isotropic the anisotropic corrections are small compared to the Ly$\alpha$ radiation pressure. Therefore, because $k_\nu$ is only weakly dependent on the scattering angle we may evaluate the energy component as $G_0^0 \approx c^{-1} \int \text{d}\nu k_\nu \int \text{d}\Omega (I_\nu - J_\nu) = 0$. The isotropic approximation also simplifies the momentum calculation: $\bmath{G}_0 \approx c^{-1} \int \text{d}\nu k_\nu \int \text{d}\Omega (I_\nu - J_\nu) \bmath{n} = c^{-1} \int \text{d}\nu k_\nu \bmath{F}_\nu$. Putting this together with the ionizing radiation gives the following:
\begin{equation} \label{eq:Monte-Carlo_estimators}
  \bmath{G}_0 = \rho (\bmath{a}_\alpha + \bmath{a}_\gamma)
  \qquad\;\; \text{and} \qquad\;\; G_0^0 = \frac{\Gamma - \Lambda}{c} \, ,
\end{equation}
where the accelerations are based on Equations~(\ref{eq:F_cell_MCE})~and~(\ref{eq:a_gamma}).

\subsection{Lagrangian radiation hydrodynamics}
We may now convert Equations~(\ref{eq:Eularian_RHD_1})-(\ref{eq:Eularian_RHD_3}) into the Lagrangian RHD framework. Derivatives in this frame are denoted by the uppercase differential operator $D(\bullet)/Dt \equiv \upartial(\bullet)/\upartial t + \bmath{v} \bmath{\cdot} \bmath{\nabla}(\bullet)$. The RHD equations are as follows:
\begin{align}
  &\frac{D\rho}{Dt} + \rho \bmath{\nabla} \bmath{\cdot} \bmath{v} = 0 \, , \\
  &\frac{D\bmath{v}}{Dt} + \frac{\bmath{\nabla}P_0}{\rho} = -\bmath{\nabla}\Phi + \bmath{a}_\alpha + \bmath{a}_\gamma \, , \\
  &\frac{D\epsilon_0}{Dt} + P_0 \frac{D(1/\rho)}{Dt} = \frac{\Gamma - \Lambda}{\rho} \, ,
\end{align}
accurate to second order in ($v/c$). The Lagrangian formulation has the advantage of substantially simplifying the numerical setup for one-dimensional problems. Our spherically symmetric hydrodynamics solver is based on the von Neumann-Richtmyer staggered mesh scheme described by \citet{Von_Neumann_1950}, \citet{Mezzacappa_1993}, and \citet{Castor_book_2004}. This algorithm has the advantage of adaptive resolution which is useful for tracking shock waves by allowing the grid to follow the motion of the gas. With this approach we must include artificial viscosity to damp numerical oscillations near shocks. We use both linear and quadratic viscosity in the references above.

\section{Simulation Results}
\label{sec:dynamical_impact}
\subsection{Models}
\subsubsection{Galaxy models}
In order to determine the dynamical impact of Ly$\alpha$ feedback on galaxy formation we explore a series of idealized model galaxies parametrized by the total mass and source luminosity. For simplicity we assume an NFW dark matter halo profile and thereby calculate the dark matter gravitational contribution analytically. Specifically, the dark matter density is $\rho_\text{DM}(r) = \rho_{\text{DM},0} R_\text{S}^3 / [r (R_\text{S} + r)^2]$ with a concentration parameter of $c_\text{NFW} \equiv R_\text{vir} / R_\text{S} \approx 5$, where $R_\text{vir}$ is the virial radius and $R_\text{S}$ is the scale radius.

The gas follows an isothermal density profile until it reaches the background IGM density. Specifically, $\rho(r) = \rho_\text{vir} (r / R_\text{vir})^{-2}$ with the normalization given by $\rho_\text{vir} = \Omega_\text{b} \Delta_\text{c}(z) H(z)^2 R_\text{vir}^2 / (8 \upi G)$, where the fractional baryon density is $\Omega_\text{b} \approx 0.0485$ and the virialization overdensity is $\Delta_\text{c}(z=10) \approx 178$. The background IGM density is $\rho_\text{IGM}(z) = \rho_{\text{crit},0} \Omega_\text{b} (1 + z)^3$, with a present-day critical density of $\rho_{\text{crit},0} \equiv 3 H_0^2 / (8 \upi G)$. The galactic gas starts off cold and neutral, with an initial temperature set by the cosmic microwave background~(CMB) at $T_\text{CMB} = 2.725~\text{K}~(1 + z)$. The exact value makes little difference as the gas quickly becomes hot and ionized within the growing \HII\ region. The simulations begin at a redshift of $z = 10$ with an initial velocity profile following the Hubble flow, i.e. $\bmath{v} = H(z) \bmath{r}$. We follow the evolution for $10$~Myr and perform the Ly$\alpha$ MCRT calculations with every third hydrodynamical timestep. In primordial gas we use an ideal gas adiabatic index of $\gamma_\text{ad} = 5/3$.

\subsubsection{Source parameters}
The spectral energy distribution from the starburst determines the ionizing radiation and Ly$\alpha$ emission which are modeled self-consistently throughout the simulation. Our code assumes a blackbody source and calculates the average emission rates, cross-sections, and heating from ionizing photons in each frequency band according to Equations~(\ref{eq:Ndot_ion})-(\ref{eq:sigma_ion}). We model the central source as consisting of $50~\Msun$ Pop III stars with an effective blackbody temperature of $T_\text{eff} = 10^{4.922}$~K and bolometric luminosity per star of $10^{5.568}~\Lsun$. If the star formation efficiency is $f_\star$ then the total luminosity of the source is
\begin{equation} \label{eq:L_star_source}
  L_\star \approx 4.5 \times 10^{42}~\text{erg\,s}^{-1}~\left( \frac{f_\star}{10^{-3}} \right) \left( \frac{f_\text{b}}{0.16} \right) \left( \frac{M_\text{vir}}{10^9~\Msun} \right) \, ,
\end{equation}
where $f_\text{b} \equiv \Omega_\text{b} / \Omega_\text{m}$ is the baryonic mass fraction. The corresponding rate of ionizing photons is related to the Ly$\alpha$ luminosity via
\begin{align} \label{eq:L_alpha_source}
  L_\alpha &\approx 0.68 h \nu_\alpha (1 - f_\text{esc}) \dot{N}_\text{ion} \notag \\
           &\approx 9.9 \times 10^{41}~\text{erg\,s}^{-1}~\left( \frac{f_\star}{10^{-3}} \right) \left( \frac{f_\text{b}}{0.16} \right) \left( \frac{M_\text{vir}}{10^9~\Msun} \right) \, ,
\end{align}
where the escape fraction of ionizing photons $f_\text{esc}$ is approximately zero in our spherically symmetric models. For reference, the Ly$\alpha$ mass-to-light ratio for this source is $\Upsilon_\alpha = M_\text{vir} / L_\alpha \approx 3.88~(f_\star/10^{-3})^{-1}$ in solar units. \hl{In Equation}~(\ref{eq:L_alpha_source}) \hl{the Ly$\alpha$ luminosity can be boosted by a factor of a few for metal-free gas illuminated by Pop III type spectra as a higher mean ionizing photon energy increases the standard conversion factor of $0.68$ from ionizing to Ly$\alpha$ photons} \citep{Raiter_2010}.

\subsubsection{Simulation parameters}
The numerical resolution is highest for the smallest mass haloes. For example, with $\approx 2000$ Lagrangian mass elements we achieve a resolution of $m_\text{cell} \approx \{164, 763, 4860, 4.8 \times 10^4, 3.8 \times 10^5 \}~\Msun$ for $M_\text{vir} = \{ 10^6, 10^7, 10^8, 10^9, 10^{10} \}~\Msun$, which typically leads to an adaptive resolution of $\approx 0.1-2$~pc at the shell front. To avoid boundary effects the computational domain is twice the virial radius, except in cases where this is smaller than a few kpc or larger than tens of kpc. The computational cost may change dramatically based on the number of timesteps with MCRT calculations and spatial resolution. To be safe we require convergence criteria such that between batches of $\approx 1600$ photons the Ly$\alpha$ force has a $\lesssim 1$ per cent relative change in $\gtrsim 99$ per cent of the cells. We bin the line of sight Ly$\alpha$ spectra with a resolution of $\Delta v \approx 1$~km\;s$^{-1}$.

  \begin{figure}
    \centering
    \includegraphics[width=\columnwidth]{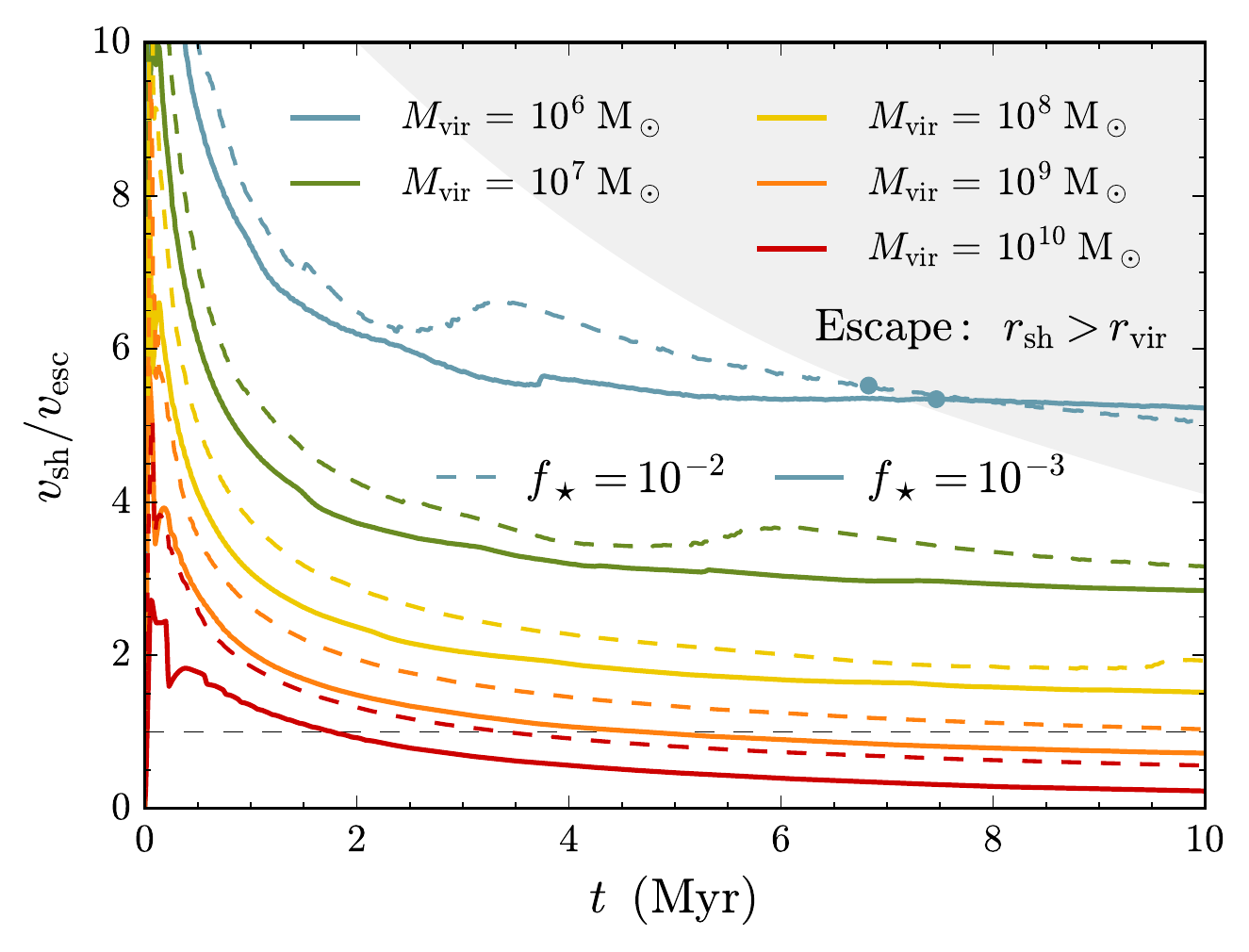}
    \caption{\protect\input{figures/v_esc/caption}}
    \label{fig:v_esc}
  \end{figure}

  \begin{figure}
    \centering
    \includegraphics[width=\columnwidth]{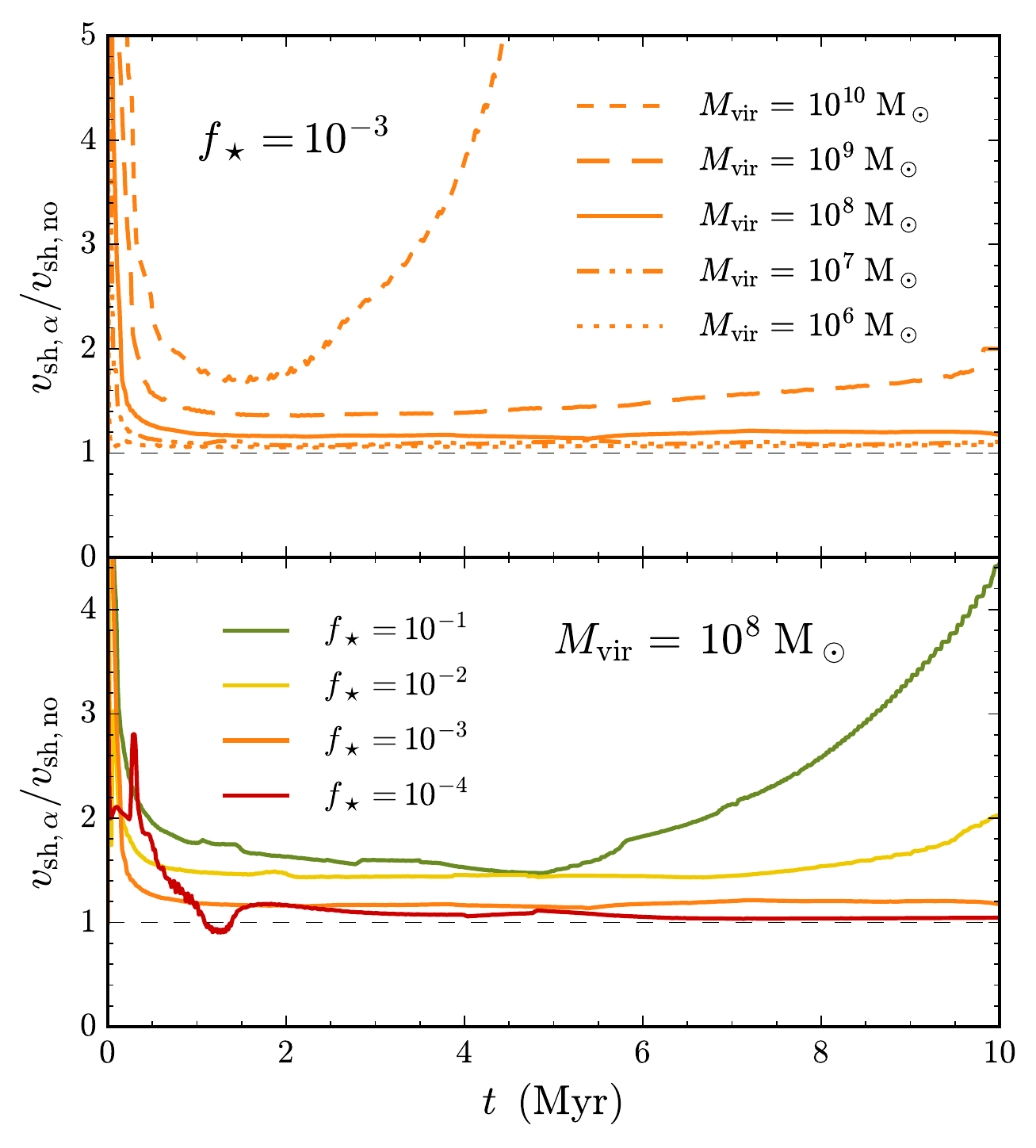}
    \caption{\protect\input{figures/v_shell_fM/caption}}
    \label{fig:v_shell_fM}
  \end{figure}

  \begin{figure}
    \centering
    \includegraphics[width=\columnwidth]{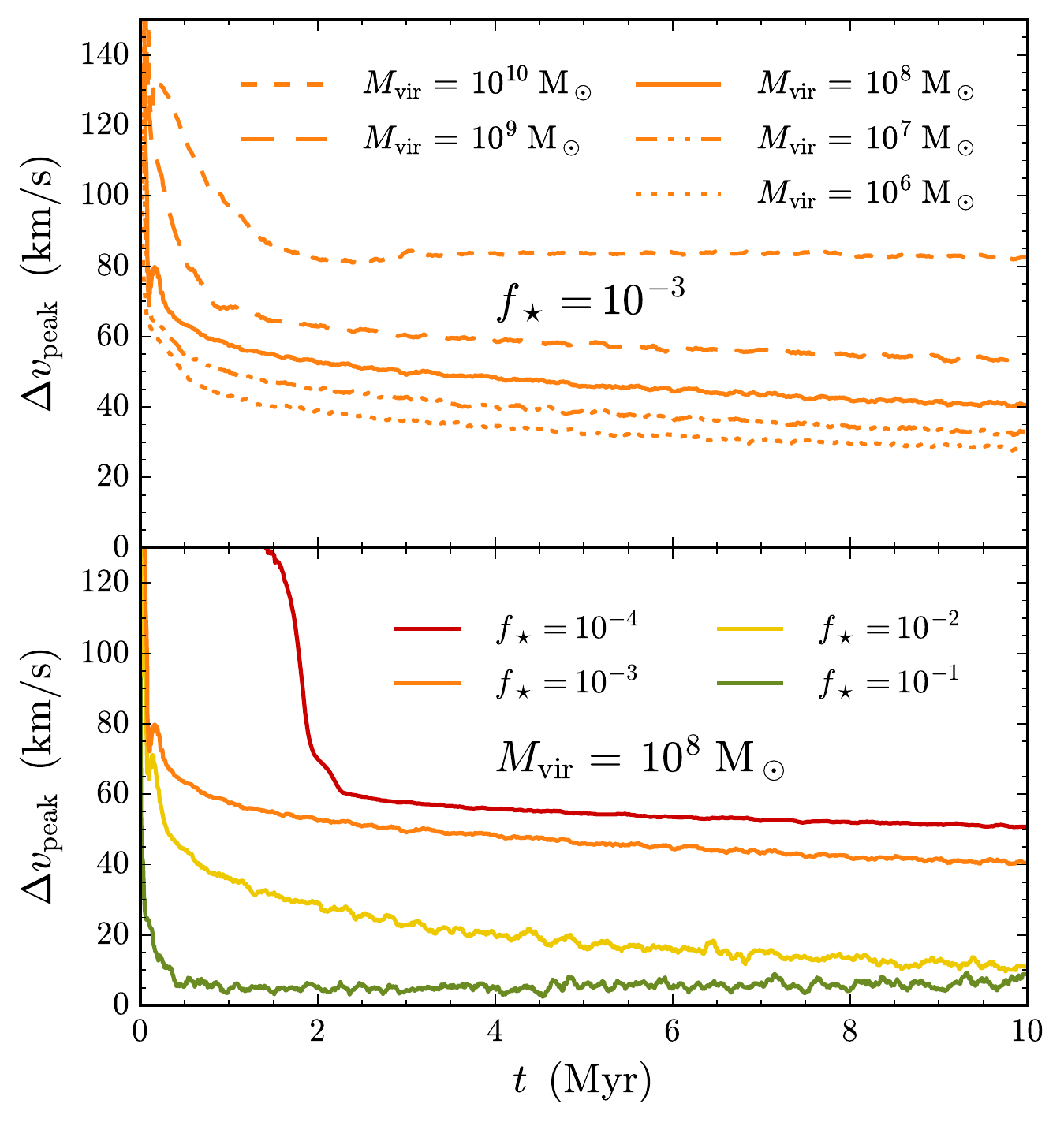}
    \caption{\protect\input{figures/v_offset_fM/caption}}
    \label{fig:v_offset_fM}
  \end{figure}

\subsection[Dynamical impact of Ly$\alpha$ radiation pressure]{Dynamical impact of Ly$\balpha$ radiation pressure}
A dense shell-like outflow structure forms in hydrodynamical response to the central starburst ionizing and heating the gas within the galaxy. The radial expansion of the shells normalized to the local escape velocity is shown for each simulation in Fig.~\ref{fig:v_esc}. The sequence of curves demonstrates the significant role of halo mass in retaining gas despite strong radiative feedback. Indeed, the shallow gravitational potential wells of minihalos allow the shell velocity to exceed the escape velocity\footnote{We note that Fig.~\ref{fig:v_esc} uses the local escape velocity, which may be lower than what is actually required to escape the galaxy. Alternatively, one might use the definition $v_\text{esc} \equiv \sqrt{2 | \Phi |}$ where the gravitational potential is given by $\Phi = G M_{<r}/r + \int_r^{r_\text{vir}} 4 \upi G r' \rho(r') \text{d}r'$. At the virial radius the second term vanishes and reduces to the definition based on the local enclosed mass.} $v_\text{esc} \equiv ( 2 G M_{<r}/r )^{1/2}$. The typical shell velocity ranges from $v_\text{sh} \approx 30-100$~km\;s$^{-1}$ for the $M_\text{vir} = 10^{6-10}~\Msun$ galaxies, which is roughly the same order of magnitude despite the large range of halo masses. This may be understood in terms of a simplified model in which the initial push from radiation pressure produces a shock propagating at a multiple of the sound speed~$c_\text{s}$. Conservation of momentum then dictates that $\frac{\text{d}}{\text{d}t}(m_\text{sh} v_\text{sh}) = \dot{p}_\text{rad} - 4 \pi r_\text{sh}^2 \rho c_\text{s}^2 - G m_\text{sh} M_{<r}/r_\text{sh}^2$, where we have included terms for radiative momentum transfer, gas pressure, and gravity. The simulations empirically motivate the adoption of the Mach number~$\mathcal{M}$ as a parametrization of the shell velocity, i.e. $v_\text{sh} \sim \mathcal{M} c_\text{s}$. The exact value of $\mathcal{M}$ is difficult to determine \textit{a priori} and implicitly depends on the hydrodynamics and halo properties, e.g. $M_\text{vir}$ and $f_\star$. However, with this in mind we calculate the ratio of the shell velocity to the escape velocity at the virial radius:
\begin{align}
  \frac{v_\text{sh}}{v_\text{esc}} & \approx \frac{\mathcal{M} c_\text{s}}{v_\text{esc}} = \mathcal{M} \left( \frac{k_\text{B} T / m_\text{H}}{G M_\text{vir} / r_\text{vir}} \right)^{1/2} \notag \\
  & \approx 1.2 \, T_4^{1/2} \left( \frac{\mathcal{M}}{5} \right) \left( \frac{M_\text{vir}}{10^9~\Msun} \right)^{-1/3} \left( \frac{1+z}{11} \right)^{-1/2} \, .
\end{align}
This quantity reflects the likelihood that galaxies retain their gas under strong radiative feedback. Qualitatively, minihalos are more susceptible to shell ejection than the more massive galaxies. We note that this simple argument is in rough agreement with the observed galaxy stellar mass function, which has a much shallower faint end slope than expected from dark matter only simulations \citep{Somerville_Dave_2015}.

In Fig.~\ref{fig:v_shell_fM} we consider the role of $M_\text{vir}$ and $f_\star$ on the relative importance of Ly$\alpha$ radiation pressure on the shell velocity compared to simulations without Ly$\alpha$ feedback. We find that for a fixed star formation efficiency including the Ly$\alpha$ force has a greater dynamical impact in larger mass haloes. We also find that for a fixed halo mass a higher star formation efficiency leads to a greater difference for simulations incorporating Ly$\alpha$ feedback. Both of these effects are likely due to a higher energy density of trapped Ly$\alpha$ photons, which leads to more scatterings because of the larger optical depth for a fixed $f_\star$ or the increased Ly$\alpha$ emission rate for a fixed $M_\text{vir}$. We find that Ly$\alpha$ radiation pressure can be dynamically important for a number of realistic protogalaxy environments.

\subsection[Predictions for Ly$\alpha$ observations]{Predictions for Ly$\balpha$ observations}
\label{sec:lya_predictions}
\begin{figure}
    \centering
    \includegraphics[width=\columnwidth]{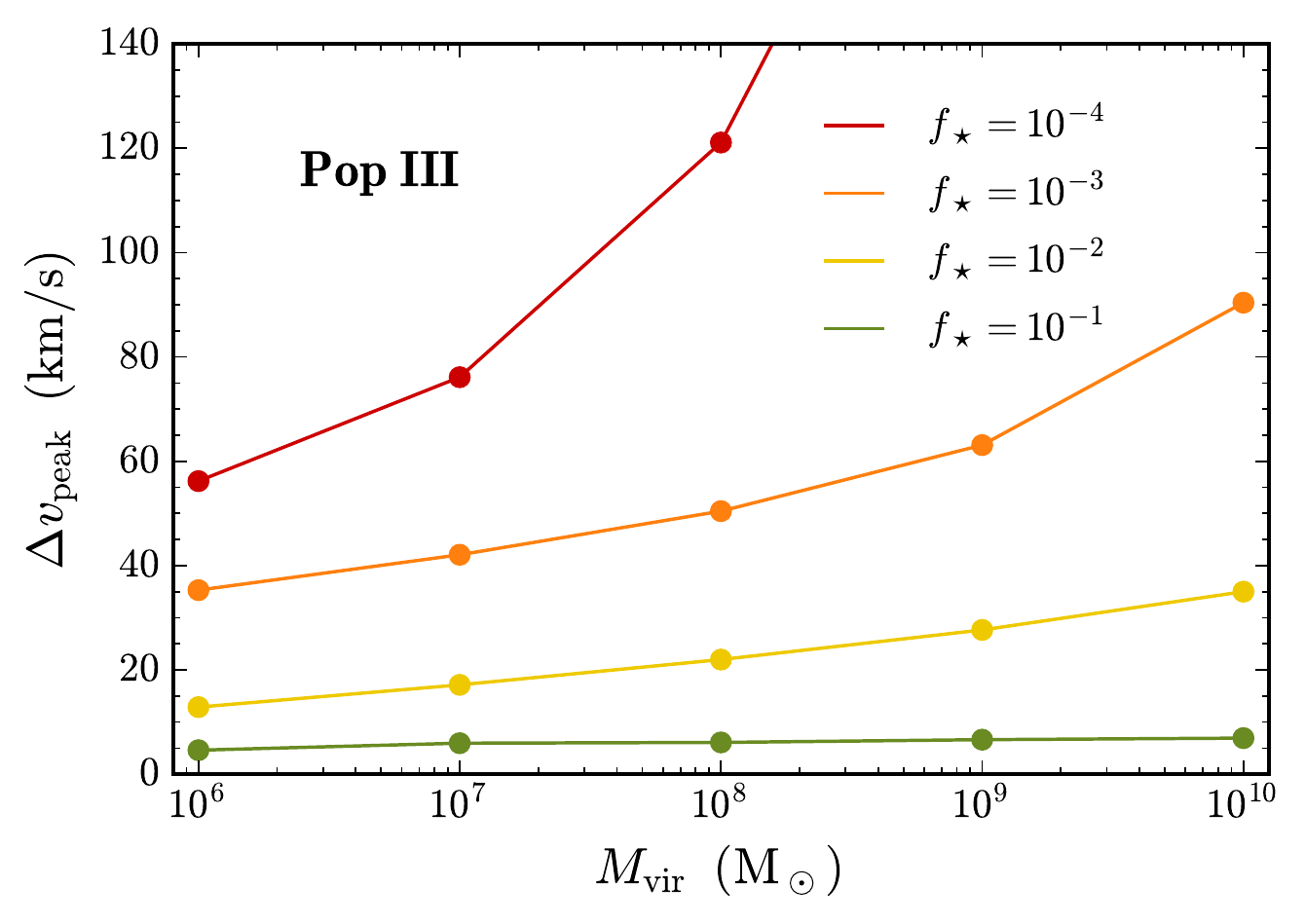}
    \caption{\protect\input{figures/v_M/caption}}
    \label{fig:v_M}
  \end{figure}

\begin{figure}
    \centering
    \includegraphics[width=\columnwidth]{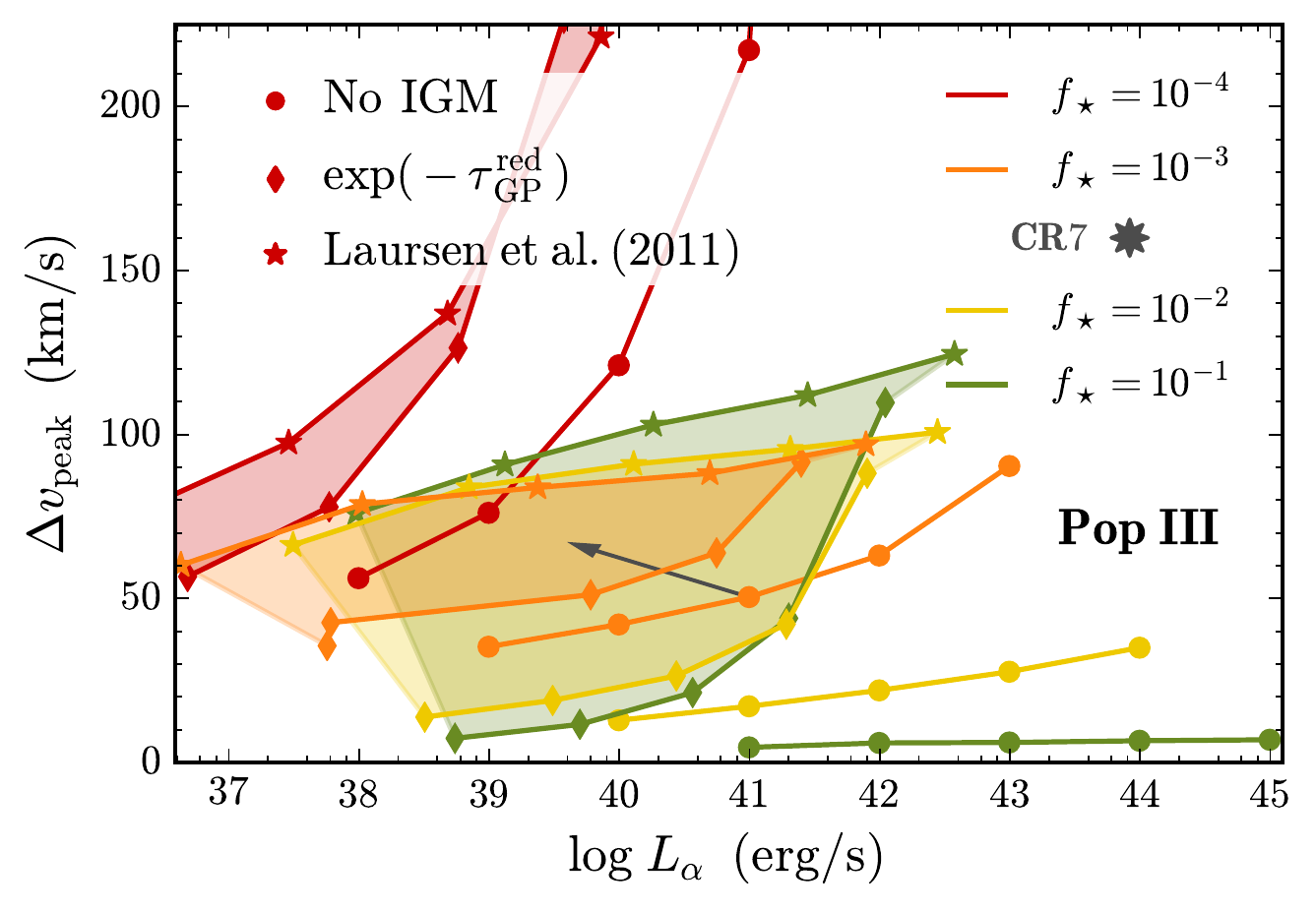}
    \caption{\protect\input{figures/v_L/caption}}
    \label{fig:v_L}
  \end{figure}

In Fig.~\ref{fig:v_offset_fM} we explore the impact of $M_\text{vir}$ and $f_\star$ on the velocity offset with respect to the central point source for the red peak of the intrinsic line-of-sight Ly$\alpha$ flux. The location of the offset is fairly constant throughout the simulations although there is some decrease in time. We find that for a fixed star formation efficiency a more massive halo has a greater velocity offset. Likewise, for a fixed halo mass a lower star formation efficiency yields a greater offset. Both of these effects are likely due to the larger optical depth for individual photons to escape the galaxy. To consolidate these trends we plot the time-averaged velocity offset as a function of halo mass in Fig.~\ref{fig:v_M}. Although there is some degeneracy at different times between these parameters the qualitative behavior is apparent. The efficiency of star formation has a strong impact on the velocity offset while the mass affects the luminosity. However, as the mass of the galaxy is not a direct observable, we also provide the velocity offsets as a function of the observed Ly$\alpha$ luminosity in Fig.~\ref{fig:v_L}. The intrinsic (circles) and observed (diamonds and stars) time-averaged velocity offsets follow similar trends as those discussed above. However, the Ly$\alpha$ luminosity that survives transmission through the neutral IGM at $z = 10$ may be significantly reduced. Furthermore, the observed velocity offset may also appear much redder than the spectra emerging from the galaxy. This overall shift from intrinsic to observed Ly$\alpha$ characterization is essential to interpreting theoretical predictions and observational bias of high-$z$ Ly$\alpha$ (non)detections.

It may be difficult to disentangle the effects of intrinsic galaxy parameters and IGM models, but larger samples of high-$z$ galaxies with measurements for $L_{\alpha,\text{obs}}$ and $\Delta v_\text{peak}$ may provide insights regarding the natures of Ly$\alpha$ emitting galaxies. For example, we expect Ly$\alpha$ detections to be biased towards bright galaxies within $\gtrsim 1$~Mpc ionized patches of the Universe or galaxies with a significant flux emerging $\gtrsim 100~$km\;s$^{-1}$ redward of the Ly$\alpha$ line centre. We include two treatments of frequency dependent transmission through the IGM in Fig.~\ref{fig:v_L} to demonstrate a range of Ly$\alpha$ reprocessing scenarios. The model labeled as \citet{Laursen_2011} is represented by stars and employs their ``benchmark'' Model~1 curve at $z \approx 6.5$ as described in relation to their figures~2~and~3. The curve is a statistical average of a large number of sightlines ($\gtrsim 10^3$) cast from several hundreds of galaxies through a simulated cosmological volume. A region of the Universe undergoing early reionization at higher redshifts may have a qualitatively similar transmission curve. The model labeled $\exp\left(-\tau_\text{GP}^\text{red}\right)$ is based on the analytic prescription of \citet{Madau_Rees_2000}, which accounts for the optical depth of the red damping wing of the Gunn-Peterson~(GP) trough. After removing all Ly$\alpha$ photons blueward of the circular velocity to account for cosmological infall \citep[see][]{Dijkstra_2007}, we attenuate the remaining Ly$\alpha$ profile based on the likelihood of undergoing a scattering event in a uniform, neutral IGM. We assume the galaxies reside within a local ionized bubble with a physical radius of $r_\text{\HII} \approx 500$~kpc and that reionization is complete by redshift $z_\text{re} \approx 6$. Any IGM model should be considered in the context of statistical variations depending on the particular galaxy and sightline. Therefore, the observables presented in this paper represent optimistic yet plausible expectations of averaged IGM effects. See \citet{Dijkstra_2014} for a review containing additional discussion and related references.

\subsection{Stellar vs. black hole source spectra}
\label{subsec:DCBH}
We now explore black hole sources as an alternative production mechanism for Ly$\alpha$ radiation pressure. We discuss this scenario in the context of so-called ``direct collapse'' black holes~(DCBHs) \citep{Bromm_Loeb_2003,Johnson_Haardt_2016}, invoked as seeds for the growth of supermassive black holes~(SMBHs) of a few billion solar masses observed in $z > 6$ quasars \citep{Fan_2006,Mortlock_2011,Wu_2015}. The DCBH model in this study employs the nonthermal Compton-thick spectrum presented by \citet{Pacucci_MBH_Spectra_2015}. The stellar and DCBH SEDs differ significantly in their effective photoionization cross-sections and energy transfer per ionizing photon as calculated according to Equations~(\ref{eq:Ndot_ion})--(\ref{eq:sigma_ion}). The Compton-thick environment reprocesses the broadband spectrum such that only X-ray and non-ionizing photons remain. This DCBH model is qualitatively different than a typical quasar spectrum, which retains the UV ionizing photons and is therefore more similar to the stellar (blackbody) source except the Ly$\alpha$ line profile is significantly broader. For example, \citet{Vanden_Berk_2001} \hl{find that the composite quasar spectrum from the Sloan Digital Sky Survey~(SDSS) has an average Ly$\alpha$ emission-line profile width of $\sigma_{\lambda,\alpha} \approx 19.46~\text{\AA} \approx 4800$~km\;s$^{-1}$.}

We present a select cases to match the most common physical conditions for DCBH formation, i.e. $M_\text{vir} \approx 10^{8-9}~\Msun$ and $M_{\bullet} \approx 10^{5-6}~\Msun$. Specifically, we use the supplementary data provided by \citet{Pacucci_MBH_Spectra_2015} to calculate typical ionization rates for their low-density profile, standard accretion scenario at three stages representing early (5~Myr), intermediate (75~Myr), and late (115~Myr) hydrodynamical response during the growth of the black hole. This is done to cover a broader range of DCBH properties, analogous to the star formation efficiency parameter. We also add an additional parameter which we call the black hole escape fraction $f_\text{esc}^{\bullet} \approx 0.01$ in order to account for three-dimensional effects. In principle this allows partial leakage of ionizing photons from the Compton-thick reprocessing region at $\lesssim 1$~pc. For our purposes this means we employ a combination of the Compton-thick spectrum and the unprocessed spectrum scaled to the mass of the black hole, i.e. $f_{\nu,\text{DCBH}} = (1-f_\text{esc}^{\bullet}) f_{\nu,\text{Compton-thick}} + f_\text{esc}^{\bullet} f_{\nu,\text{unprocessed}}$. We obtain the Ly$\alpha$ and \HeII\ line luminosities by integrating the \citet{Pacucci_MBH_Spectra_2015} spectra around the 1216~\AA\ and 1640~\AA\ peaks, respectively. For concreteness, we provide a list of DCBH source properties for each model in Table~\ref{tab:DCBH}. Even a small fraction of escaped, unprocessed ionizing photons has an impact on the ionization properties. The effective cross-sections and energy transferred per photon are not very sensitive to the different stages, so we only provide representative values for Stage~II as follows: $\langle\sigma_\text{\HI}\rangle_{1-3} = \{3.1 \times 10^{-18}, 4.99 \times 10^{-19}, 3.42 \times 10^{-20}\}~\text{cm}^2$, $\langle\sigma_\text{\HeI}\rangle_{2-3} = \{4.3 \times 10^{-18}, 7.62 \times 10^{-19}\}~\text{cm}^2$, $\langle\sigma_\text{\HeII}\rangle_3 = 4.72 \times 10^{-19}~\text{cm}^2$, $\langle\varepsilon_\text{\HI}\rangle_{1-3} = \{3.65, 18.7, 58.1\}$~eV, $\langle\varepsilon_\text{\HeI}\rangle_{2-3} = \{9.58, 54.4\}$~eV, and $\langle\varepsilon_\text{\HeII}\rangle_3 = 18.7$~eV. Note, for simplicity we model the cascade of multiple ionization and heating events experienced by the X-ray photons with a `one-shot' approximation, where all the energy is transferred in a single scattering \citep{Shull_1985}.

\begin{table}
  \caption{Summary of the source parameters for the DCBH models. The stages represent early (5~Myr), intermediate (75~Myr), and late (115~Myr) hydrodynamical response during the growth of the black hole. The black hole escape fraction $f_\text{esc}^{\bullet} \approx 0.01$ accounts for three-dimensional effects allowing some of the ionizing photons to escape without significant reprocessing.}
  \label{tab:DCBH}
  \begin{tabular}{@{} c cccc @{}}
    \hline
    Model & Stage I & Stage II & Stage III \\
    \hline
    \vspace{.05cm}
    \;$t_{\bullet}$ \; [Myr]\; & 5 & 75 & 115 \\
    \;$M_{\bullet}$ \; [$\Msun$]\;            & $1.33 \times 10^5$    & $6.52 \times 10^5$    & $1.48 \times 10^6$ \\
    \;$L_\alpha$ \;[erg\,$s^{-1}$]\;          & $8.72 \times 10^{42}$ & $3.63 \times 10^{43}$ & $6.43 \times 10^{43}$ \\
    \;$L_\text{\HeII}$ \;[erg\,$s^{-1}$]\;    & $1.54 \times 10^{42}$ & $6.76 \times 10^{42}$ & $5.75 \times 10^{42}$ \\
    \;$\dot{N}_{\text{ion},1}$ \;[s$^{-1}$]\; & $1.13 \times 10^{51}$ & $5.52 \times 10^{51}$ & $1.26 \times 10^{52}$ \\
    \;$\dot{N}_{\text{ion},2}$ \;[s$^{-1}$]\; & $1.53 \times 10^{51}$ & $7.52 \times 10^{51}$ & $1.71 \times 10^{52}$ \\
    \;$\dot{N}_{\text{ion},3}$ \;[s$^{-1}$]\; & $1.53 \times 10^{51}$ & $7.95 \times 10^{51}$ & $2.60 \times 10^{52}$ \\
    \hline
  \end{tabular}
\end{table}

  \begin{figure}
    \centering
    \includegraphics[width=\columnwidth]{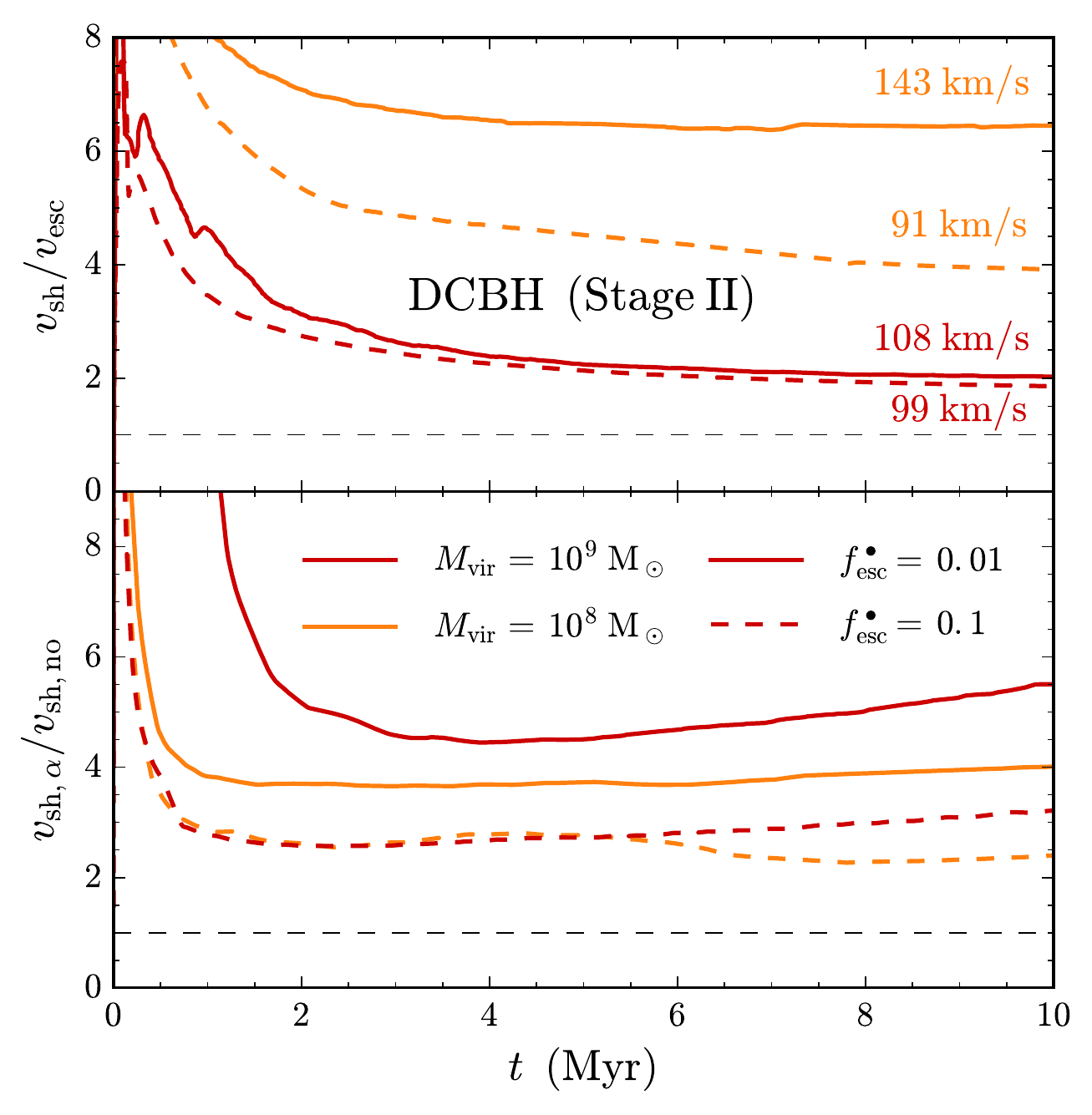}
    \caption{\protect\input{figures/v_shell_DCBH/caption}}
    \label{fig:v_shell_DCBH}
  \end{figure}

\begin{figure}
    \centering
    \includegraphics[width=\columnwidth]{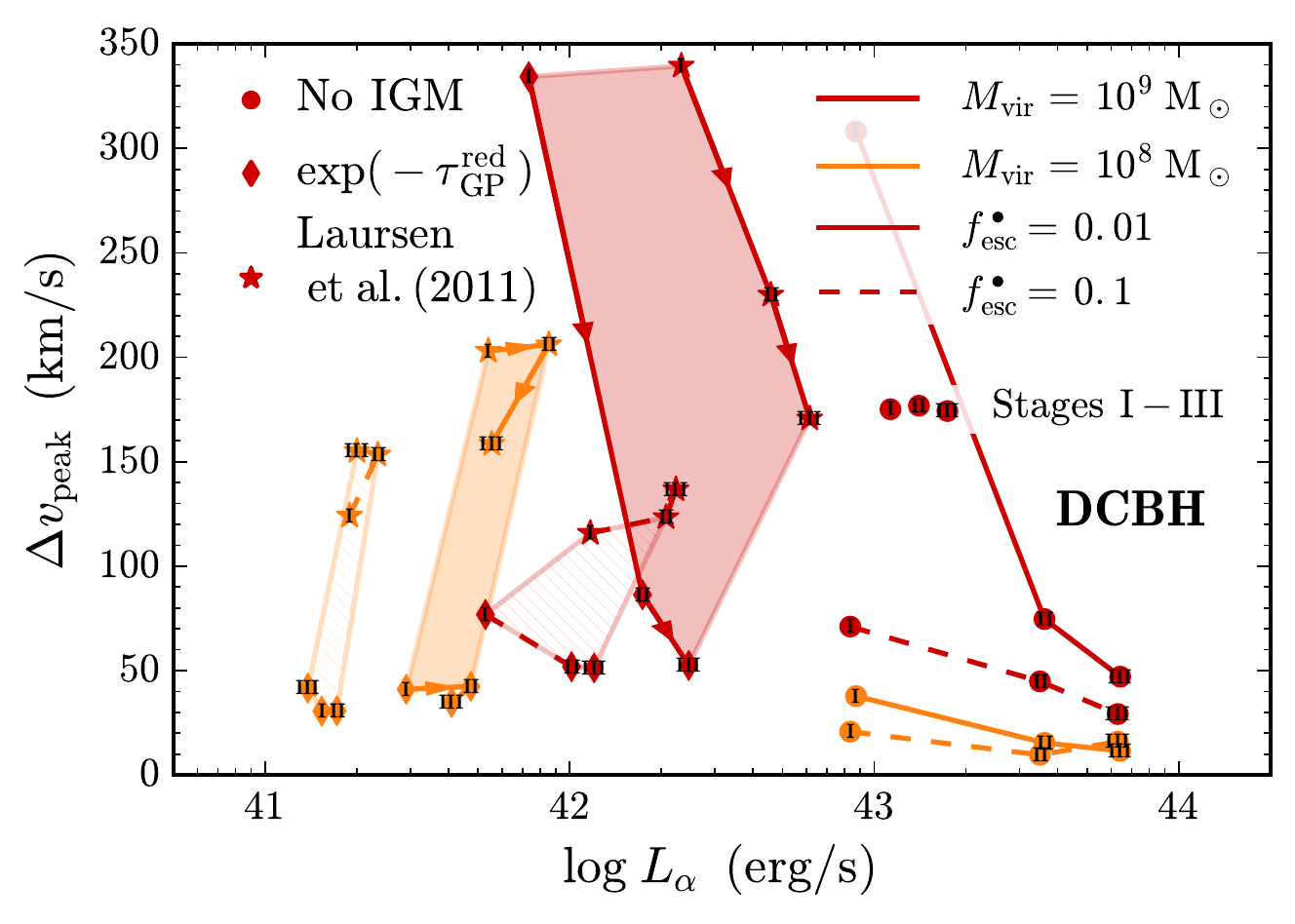}
    \caption{\protect\input{figures/v_L_DCBH/caption}}
    \label{fig:v_L_DCBH}
  \end{figure}

As shown in Fig.~\ref{fig:v_shell_DCBH} the DCBH scenario is an ideal setting for Ly$\alpha$ radiation pressure to have a significant dynamical impact in primordial galaxies. In every case Ly$\alpha$ feedback increases the simulated shell velocity by at least a factor of a few. Ly$\alpha$ radiation pressure has a greater relative impact for larger haloes and lower black hole escape fractions. Furthermore, this allows the shell to exceed the escape velocity throughout the simulations, even for the $M_\text{vir} = 10^9~\Msun$ halo. Finally, in Fig.~\ref{fig:v_L_DCBH} we provide the time-averaged velocity offsets from the Compton-thick black hole source as a function of the observed Ly$\alpha$ luminosity. The intrinsic and observed points are similar to the stellar case in Fig.~\ref{fig:v_L} except the velocity offsets are generally larger. This is likely due to the higher residual \HI\ fraction resulting from the harder spectrum. \hl{Figures}~\ref{fig:v_L}~\hl{and}~\ref{fig:v_L_DCBH} \hl{both use a simplified IGM model (diamond markers) that assumes a neutral IGM outside a local} \HII\ \hl{superbubble. However, the impact of the IGM weakens when reionization is underway and by $z \sim 10$ the evolving, inhomogeneous ionization state can enhance the overall transmission of Ly$\alpha$ photons through the IGM. Furthermore, the analytic IGM model suppresses the Ly$\alpha$ flux by more than one order of magnitude. However, even when the Universe is still 90\% neutral by volume, patchy reionization allows the IGM to transmit $\gtrsim 10$\% of the flux} \citep[see figures 2 and 3 of][]{Dijkstra_2011}. \hl{This transmission applies to haloes more massive than considered in this paper, which are more likely to reside in larger ionized bubbles. Still, their calculation is relevant because both the DCBH and Pop~III scenarios may benefit from having a more massive galaxy nearby, either to provide the critical Lyman-Werner flux or to keep the halo from forming stars until it crosses $T_\text{vir} \sim 10^4$~K} \citep[see e.g.][]{Visbal_CR7_2016}.

\section{Summary and Conclusions}
\label{sec:conc}
Ly$\alpha$ radiative transfer simulations not only provide insight about observations of Ly$\alpha$ emitting galaxies but also regarding fundamental processes affecting the gas within the galaxies. As the observational frontier for high-$z$ galaxies extends into the pre-reionization Universe the conditions for which Ly$\alpha$ radiation pressure may be dynamically important are increasingly common. Therefore, we have developed a one-dimensional radiation hydrodynamics framework for radiatively-driven outflows during the formation of the first galaxies. Our simulations represent the first hydrodynamical study incorporating accurate Monte-Carlo radiative transfer calculations of Ly$\alpha$ feedback. This is essential because order of magnitude estimates and even post-processing simulations are unlikely to be self-consistent with the gas and ionization dynamics. Still, these calculations indicate that multiple scattering within high \HI\ column density shells is capable of significantly enhancing the effective Ly$\alpha$ force \citep{Dijkstra_Loeb_2008,Dijkstra_Loeb_2009}. Furthermore, Ly$\alpha$ radiation pressure may contribute alongside other feedback mechanisms such as supernova explosions. Understanding the role of Ly$\alpha$ trapping in galaxies is a challenging but important problem.

We have studied Ly$\alpha$ radiation pressure in the context of galaxies at $z \approx 10$ with different halo mass $M_\text{vir}$ and star formation efficiency $f_\star$ defined as the fraction of baryonic mass that is in stars. We have also considered Ly$\alpha$ feedback within the context of DCBH formation. The central starburst or black hole emission drives an expanding shell of gas from the centre. The results of our one-dimensional models may be summarized as follows:
\begin{itemize}
  \item[(1)] Strong radiative feedback in minihaloes can launch supersonic outflows into the IGM and Ly$\alpha$ radiation pressure may enhance this effect,
  \item[(2)] Including Ly$\alpha$ feedback has a greater relative impact on the shell velocity with higher $f_\star$ or larger $M_\text{vir}$,
  \item[(3)] The velocity offset $\Delta v_\text{peak}$ of the Ly$\alpha$ flux depends on the \HI\ optical depth, therefore a higher neutral fraction due to a lower $f_\star$ or larger $M_\text{vir}$ translates to a greater velocity offset,
  \item[(4)] \hl{We provide quantitative estimates for the extent to which} the scattering of Ly$\alpha$ photons in the IGM leads to a reprocessing of the observed flux, i.e. the observed Ly$\alpha$ luminosity is reduced and the velocity offset undergoes a redward shift,
  \item[(5)] Ly$\alpha$ radiation pressure may have a significant dynamical impact on gas surrounding DCBHs, with Ly$\alpha$ signatures typically characterized by larger velocity offsets than stellar counterparts \hl{if in both cases the line shift is set by Ly$\alpha$ radiation pressure}.
\end{itemize}
We emphasize that the details of these conclusions may rely on particular modeling choices. Still, the main results are fairly robust. The most significant uncertainties are likely associated with the one-dimensional approximation, which provides broad insights but three-dimensional effects may be important. Post-processing studies of Ly$\alpha$ radiative transfer in similar contexts inform us about the impact of three-dimensional effects \citep{Dijkstra_Kramer_2012,Behrens_2014,Duval_2014,Zheng_Wallace_2014,Smith_2015}. For example, geometry, gas clumping, rotation, filamentary structure, and anisotropic emission from the source often lead to anisotropic escape, photon leakage, or otherwise altered dynamical impact. Ly$\alpha$ observables such as the equivalent width and escape fraction may also be affected. For additional discussion regarding the caveats of these models and methods see \citet{Smith_CR7_2016}, which is also useful when applying these results to observations of high-$z$ Ly$\alpha$ emitting galaxies.

We also note that there are other ways to trigger outflows which may lead to different results for the Ly$\alpha$ signatures. In star-forming galaxies supernova explosions could lead to even faster winds leading to larger Ly$\alpha$ velocity offsets until the velocity gradient facilitates escape from Doppler shifting. Additionally, these environments are likely to already be metal enriched such that the Ly$\alpha$ luminosity could be several times fainter due to dust absorption. However, some of these uncertainties could be disentangled by comparing the Ly$\alpha$ properties with complementary multi-frequency observations. Additionally, our theoretical framework relies on the ability to measure the velocity offset of high-$z$ galaxies, which is only possible if there is a nonshifted line to compare with. If the \HeII\ 1640~\AA\ line is not observed then it may still be possible to observe Balmer lines. \hl{The \textit{JWST} is capable of detecting the H$\alpha$, H$\beta$, H$\gamma$, and H$\delta$ lines with the Near-Infrared Spectrograph (NIRSpec) at a medium spectral resolution of $R = 2700$ out to redshifts of $z = 6.6$, $9.3$, $10.5$, and $11.2$ respectively. Beyond that the lines fall into the wavelength coverage of the Mid-Infrared Instrument (MIRI), which is an order of magnitude less sensitive than the NIRSpec for line flux detection at comparable spectral resolution. Even though the ratio of Balmer-line intensities implies diminishing returns, i.e. the Balmer decrement is typically H$\alpha$:H$\beta$:H$\gamma$ = 2.86:1:0.47} \citep{Osterbrock_book_2006}, \hl{depending on the source redshift it may still be advantageous to look for a higher-order line, considering the relative spectroscopic performance gain of NIRSpec over MIRI.} Therefore, we expect selected samples of high-$z$ Ly$\alpha$ emitters will also include the measured velocity offset similar to current lower-redshift surveys.

Recently, the luminous COSMOS redshift~7~(CR7) Ly$\alpha$ emitter at $z = 6.6$ was confirmed to have strong \HeII\ emission with no detection of metal lines from the UV to the near infrared \citep{Matthee_2015,Sobral_2015}. As a result several groups have considered the CR7 source in the context of a young primordial starburst or direct collapse black hole \citep{Pallottini_2015,Agarwal_2016,Hartwig_2015,Visbal_CR7_2016,Dijkstra_DCBH_2016,Smidt_CR7_2016}. CR7 and similar galaxies therefore represent an ideal application of the methodology presented herein as they allow for direct comparison with current and upcoming observations. In \citet{Smith_CR7_2016}, we closely examined and reproduced several Ly$\alpha$ signatures of the CR7 source under a DCBH model, including the velocity offset between the Ly$\alpha$ and \HeII\ line peaks. We also found that Ly$\alpha$ radiation pressure turns out to be dynamically important in the case of CR7. In the near future, other sources similar to CR7 may provide additional constraints on early galaxy and quasar formation. Indeed, \citet{Pacucci_DCBH_2016} identified two objects characterized by very red colours and robust X-ray detections in the CANDLES/GOODS-S survey with photometric redshift $z \gtrsim 6$ representing promising black hole seed candidates.

The main goal is to assemble a broad net of models to provide a theoretical framework for Ly$\alpha$ observations of high-$z$ galaxies, leading to a more complete understanding of the environments that are dynamically impacted by Ly$\alpha$ radiation pressure. The extent of Ly$\alpha$ trapping is a complex problem affected by a number of factors, including destruction via collisional (de)excitation, absorption by dust, holes or pathways for escape, reduced opacity due to turbulence and bulk gas motion, sources with broad or offset line emission, or cold gas accretion along filaments. Ly$\alpha$ radiation hydrodynamics simulations are now computationally feasible and will provide additional insights into the process of galaxy formation. Overall, in the early Universe, the higher neutral fraction acts to reduce the visibility of Ly$\alpha$ emission, implying an evolving luminosity function, as suggested by current observations. On the other hand, the same conditions favor the dynamical impact of Ly$\alpha$ radiation pressure inside the first galaxies, thus rendering radiation hydrodynamical studies crucial to fully elucidate the epoch of cosmic dawn.

\section*{Acknowledgements}
\hl{We thank the referee for an exceptional review and constructive comments which have improved the quality of this work.} AS thanks the Institute for Theory and Computation (ITC) at the Harvard-Smithsonian Center for Astrophysics for hosting him as a visitor through the National Science Foundation Graduate Research Internship Program. This material is based upon work supported by a National Science Foundation Graduate Research Fellowship. VB acknowledges support from NSF grant AST-1413501, and AL from NSF grant AST-1312034. The authors acknowledge the Texas Advanced Computing Center~(TACC) at the University of Texas at Austin for providing HPC resources.


\bibliographystyle{mnras}
\bibliography{biblio}

\begin{thebibliography}{}
\makeatletter
\relax
\def\mn@urlcharsother{\let\do\@makeother \do\$\do\&\do\#\do\^\do\_\do\%\do\~}
\def\mn@doi{\begingroup\mn@urlcharsother \@ifnextchar [ {\mn@doi@}
  {\mn@doi@[]}}
\def\mn@doi@[#1]#2{\def\@tempa{#1}\ifx\@tempa\@empty \href
  {http://dx.doi.org/#2} {doi:#2}\else \href {http://dx.doi.org/#2} {#1}\fi
  \endgroup}
\def\mn@eprint#1#2{\mn@eprint@#1:#2::\@nil}
\def\mn@eprint@arXiv#1{\href {http://arxiv.org/abs/#1} {{\tt arXiv:#1}}}
\def\mn@eprint@dblp#1{\href {http://dblp.uni-trier.de/rec/bibtex/#1.xml}
  {dblp:#1}}
\def\mn@eprint@#1:#2:#3:#4\@nil{\def\@tempa {#1}\def\@tempb {#2}\def\@tempc
  {#3}\ifx \@tempc \@empty \let \@tempc \@tempb \let \@tempb \@tempa \fi \ifx
  \@tempb \@empty \def\@tempb {arXiv}\fi \@ifundefined
  {mn@eprint@\@tempb}{\@tempb:\@tempc}{\expandafter \expandafter \csname
  mn@eprint@\@tempb\endcsname \expandafter{\@tempc}}}

\bibitem[\protect\citeauthoryear{Abdikamalov, Burrows, Ott, L\"{o}ffler,
  O'Connor, Dolence  \& Schnetter}{Abdikamalov et~al.}{2012}]{Abdikamalov_2012}
Abdikamalov E.,  Burrows A.,  Ott C.~D.,  L\"{o}ffler F.,  O'Connor E.,
  Dolence J.~C.,   Schnetter E.,  2012, \mn@doi [ApJ]
  {10.1088/0004-637X/755/2/111}, 755, 111

\bibitem[\protect\citeauthoryear{Abel, Norman  \& Madau}{Abel
  et~al.}{1999}]{Abel_1999}
Abel T.,  Norman M.~L.,   Madau P.,  1999, \mn@doi [ApJ] {10.1086/307739}, 523,
  66

\bibitem[\protect\citeauthoryear{Adams}{Adams}{1971}]{Adams_1971}
Adams T.~F.,  1971, \mn@doi [ApJ] {10.1086/151111}, 168, 575

\bibitem[\protect\citeauthoryear{Adams}{Adams}{1972}]{Adams_1972}
Adams T.~F.,  1972, \mn@doi [ApJ] {10.1086/151503}, 174, 439

\bibitem[\protect\citeauthoryear{Adams}{Adams}{1975}]{Adams_1975}
Adams T.~F.,  1975, \mn@doi [ApJ] {10.1086/153891}, 201, 350

\bibitem[\protect\citeauthoryear{Agarwal, Johnson, Zackrisson, Labbe, van~den
  Bosch, Natarajan  \& Khochfar}{Agarwal et~al.}{2016}]{Agarwal_2016}
Agarwal B.,  Johnson J.~L.,  Zackrisson E.,  Labbe I.,  van~den Bosch F.~C.,
  Natarajan P.,   Khochfar S.,  2016, \mn@doi [MNRAS] {10.1093/mnras/stw1173},
  460, 4003

\bibitem[\protect\citeauthoryear{Ahn}{Ahn}{2004}]{Ahn_2004}
Ahn S.-H.,  2004, \mn@doi [ApJ] {10.1086/381750}, 601, L25

\bibitem[\protect\citeauthoryear{Ahn, Lee  \& Lee}{Ahn et~al.}{2002}]{Ahn_2002}
Ahn S.-H.,  Lee H.-W.,   Lee H.~M.,  2002, \mn@doi [ApJ] {10.1086/338497}, 567,
  922

\bibitem[\protect\citeauthoryear{Ambarzumian}{Ambarzumian}{1932}]{Ambarzumian_1932}
Ambarzumian V.~A.,  1932, MNRAS, 93, 50

\bibitem[\protect\citeauthoryear{Behrens, Dijkstra  \& Niemeyer}{Behrens
  et~al.}{2014}]{Behrens_2014}
Behrens C.,  Dijkstra M.,   Niemeyer J.~C.,  2014, \mn@doi [A\&A]
  {10.1051/0004-6361/201322949}, 563, A77

\bibitem[\protect\citeauthoryear{Bithell}{Bithell}{1990}]{Bithell_1990}
Bithell M.,  1990, MNRAS, 244, 738

\bibitem[\protect\citeauthoryear{Bouwens et~al.,}{Bouwens
  et~al.}{2011}]{Bouwens_2011}
Bouwens R.~J.,  et~al., 2011, \mn@doi [ApJ] {10.1088/0004-637X/737/2/90}, 737,
  90

\bibitem[\protect\citeauthoryear{Bromm \& Loeb}{Bromm \&
  Loeb}{2003}]{Bromm_Loeb_2003}
Bromm V.,  Loeb A.,  2003, \mn@doi [ApJ] {10.1086/377529}, 596, 34

\bibitem[\protect\citeauthoryear{Bromm \& Yoshida}{Bromm \&
  Yoshida}{2011}]{Bromm_Yoshida_2011}
Bromm V.,  Yoshida N.,  2011, \mn@doi [ARA\&A]
  {10.1146/annurev-astro-081710-102608}, 49, 373

\bibitem[\protect\citeauthoryear{Bromm, Coppi  \& Larson}{Bromm
  et~al.}{2002}]{Bromm_2002}
Bromm V.,  Coppi P.~S.,   Larson R.~B.,  2002, \mn@doi [ApJ] {10.1086/323947},
  564, 23

\bibitem[\protect\citeauthoryear{Castor}{Castor}{2004}]{Castor_book_2004}
Castor J.~I.,  2004, {Radiation Hydrodynamics}.
Cambridge Univ. Press, Cambridge, UK

\bibitem[\protect\citeauthoryear{Cen}{Cen}{1992}]{Cen_1992}
Cen R.,  1992, \mn@doi [ApJS] {10.1086/191630}, 78, 341

\bibitem[\protect\citeauthoryear{Chandrasekhar}{Chandrasekhar}{1945}]{Chandrasekhar_1945}
Chandrasekhar S.,  1945, \mn@doi [ApJ] {10.1086/144771}, 102, 402

\bibitem[\protect\citeauthoryear{Chuzhoy \& Zheng}{Chuzhoy \&
  Zheng}{2007}]{Chuzhoy_2007}
Chuzhoy L.,  Zheng Z.,  2007, \mn@doi [ApJ] {10.1086/522491}, 670, 912

\bibitem[\protect\citeauthoryear{Cox}{Cox}{1985}]{Cox_1985}
Cox D.~P.,  1985, \mn@doi [ApJ] {10.1086/162812}, 288, 465

\bibitem[\protect\citeauthoryear{Dijkstra}{Dijkstra}{2014}]{Dijkstra_2014}
Dijkstra M.,  2014, \mn@doi [Publ. Astron. Soc. Australia]
  {10.1017/pasa.2014.33}, 31, e040

\bibitem[\protect\citeauthoryear{Dijkstra \& Kramer}{Dijkstra \&
  Kramer}{2012}]{Dijkstra_Kramer_2012}
Dijkstra M.,  Kramer R.,  2012, \mn@doi [MNRAS]
  {10.1111/j.1365-2966.2012.21131.x}, 424, 1672

\bibitem[\protect\citeauthoryear{Dijkstra \& Loeb}{Dijkstra \&
  Loeb}{2008}]{Dijkstra_Loeb_2008}
Dijkstra M.,  Loeb A.,  2008, \mn@doi [MNRAS]
  {10.1111/j.1365-2966.2008.13920.x}, 391, 457

\bibitem[\protect\citeauthoryear{Dijkstra \& Loeb}{Dijkstra \&
  Loeb}{2009}]{Dijkstra_Loeb_2009}
Dijkstra M.,  Loeb A.,  2009, \mn@doi [MNRAS]
  {10.1111/j.1365-2966.2009.14602.x}, 396, 377

\bibitem[\protect\citeauthoryear{Dijkstra, Haiman  \& Spaans}{Dijkstra
  et~al.}{2006}]{Dijkstra_2006}
Dijkstra M.,  Haiman Z.,   Spaans M.,  2006, \mn@doi [ApJ] {10.1086/506243},
  649, 14

\bibitem[\protect\citeauthoryear{Dijkstra, Lidz  \& Wyithe}{Dijkstra
  et~al.}{2007}]{Dijkstra_2007}
Dijkstra M.,  Lidz A.,   Wyithe J. S.~B.,  2007, \mn@doi [MNRAS]
  {10.1111/j.1365-2966.2007.11666.x}, 377, 1175

\bibitem[\protect\citeauthoryear{Dijkstra, Haiman, Mesinger  \&
  Wyithe}{Dijkstra et~al.}{2008}]{Dijkstra_Haiman_2008}
Dijkstra M.,  Haiman Z.,  Mesinger A.,   Wyithe J. S.~B.,  2008, \mn@doi
  [MNRAS] {10.1111/j.1365-2966.2008.14031.x}, 391, 1961

\bibitem[\protect\citeauthoryear{Dijkstra, Mesinger  \& Wyithe}{Dijkstra
  et~al.}{2011}]{Dijkstra_2011}
Dijkstra M.,  Mesinger A.,   Wyithe J. S.~B.,  2011, \mn@doi [MNRAS]
  {10.1111/j.1365-2966.2011.18530.x}, 414, 2139

\bibitem[\protect\citeauthoryear{Dijkstra, Gronke  \& Sobral}{Dijkstra
  et~al.}{2016}]{Dijkstra_DCBH_2016}
Dijkstra M.,  Gronke M.,   Sobral D.,  2016, \mn@doi [ApJ]
  {10.3847/0004-637X/823/2/74}, 823, 74

\bibitem[\protect\citeauthoryear{Duval, Schaerer, {\"{O}}stlin  \&
  Laursen}{Duval et~al.}{2014}]{Duval_2014}
Duval F.,  Schaerer D.,  {\"{O}}stlin G.,   Laursen P.,  2014, \mn@doi [A\&A]
  {10.1051/0004-6361/201220455}, 562, A52

\bibitem[\protect\citeauthoryear{Faisst}{Faisst}{2016}]{Faisst_2016}
Faisst A.~L.,  2016, ApJ, preprint
  (\href{http://arxiv.org/abs/1605.06507}{arXiv:1605.06507})

\bibitem[\protect\citeauthoryear{Fan et~al.,}{Fan et~al.}{2006}]{Fan_2006}
Fan X.,  et~al., 2006, \mn@doi [AJ] {10.1086/500296}, 131, 1203

\bibitem[\protect\citeauthoryear{Ferrara \& Loeb}{Ferrara \&
  Loeb}{2013}]{Ferrara_Loeb_2013}
Ferrara A.,  Loeb A.,  2013, \mn@doi [MNRAS] {10.1093/mnras/stt381}, 431, 2826

\bibitem[\protect\citeauthoryear{Field}{Field}{1959}]{Field_1959}
Field G.~B.,  1959, \mn@doi [ApJ] {10.1086/146654}, 129, 551

\bibitem[\protect\citeauthoryear{Finkelstein et~al.,}{Finkelstein
  et~al.}{2013}]{Finkelstein_2013}
Finkelstein S.~L.,  et~al., 2013, \mn@doi [Nature] {10.1038/nature12657}, 502,
  524

\bibitem[\protect\citeauthoryear{Gardner et~al.,}{Gardner
  et~al.}{2006}]{Gardner_2006}
Gardner J.~P.,  et~al., 2006, \mn@doi [Space Sci. Rev.]
  {10.1007/s11214-006-8315-7}, 123, 485

\bibitem[\protect\citeauthoryear{Gnedin, Kravtsov  \& Chen}{Gnedin
  et~al.}{2008}]{Gnedin_2008}
Gnedin N.~Y.,  Kravtsov A.~V.,   Chen H.-W.,  2008, \mn@doi [ApJ]
  {10.1086/524007}, 672, 765

\bibitem[\protect\citeauthoryear{Greif, Johnson, Klessen  \& Bromm}{Greif
  et~al.}{2009}]{Greif_2009}
Greif T.~H.,  Johnson J.~L.,  Klessen R.~S.,   Bromm V.,  2009, \mn@doi [MNRAS]
  {10.1111/j.1365-2966.2009.15336.x}, 399, 639

\bibitem[\protect\citeauthoryear{Gronke \& Dijkstra}{Gronke \&
  Dijkstra}{2016}]{Gronke_Multiphase_2016}
Gronke M.,  Dijkstra M.,  2016, \mn@doi [ApJ] {10.3847/0004-637X/826/1/14},
  826, 14

\bibitem[\protect\citeauthoryear{Gronke, Bull  \& Dijkstra}{Gronke
  et~al.}{2015}]{Gronke_2015}
Gronke M.,  Bull P.,   Dijkstra M.,  2015, ApJ, 812, 123

\bibitem[\protect\citeauthoryear{Haehnelt}{Haehnelt}{1995}]{Haehnelt_1995}
Haehnelt M.~G.,  1995, MNRAS, 273, 249

\bibitem[\protect\citeauthoryear{Harries}{Harries}{2015}]{Harries_2015}
Harries T.~J.,  2015, \mn@doi [MNRAS] {10.1093/mnras/stv158}, 448, 3156

\bibitem[\protect\citeauthoryear{Harrington}{Harrington}{1973}]{Harrington_1973}
Harrington J.~P.,  1973, \mn@doi [MNRAS] {10.1093/mnras/162.1.43}, 162, 43

\bibitem[\protect\citeauthoryear{Hartwig et~al.,}{Hartwig
  et~al.}{2015}]{Hartwig_2015}
Hartwig T.,  et~al., 2015, MNRAS, preprint
  (\href{http://arxiv.org/abs/1512.01111}{arXiv:1512.01111})

\bibitem[\protect\citeauthoryear{Henney \& Arthur}{Henney \&
  Arthur}{1998}]{Henney_1998}
Henney W.~J.,  Arthur S.~J.,  1998, \mn@doi [AJ] {10.1086/300433}, 116, 322

\bibitem[\protect\citeauthoryear{Jeon, Pawlik, Bromm  \&
  Milosavljevi{\'{c}}}{Jeon et~al.}{2014}]{Jeon_XRB_2014}
Jeon M.,  Pawlik A.~H.,  Bromm V.,   Milosavljevi{\'{c}} M.,  2014, \mn@doi
  [MNRAS] {10.1093/mnras/stu444}, 440, 3778

\bibitem[\protect\citeauthoryear{Jeon, Bromm, Pawlik  \&
  Milosavljevi\'{c}}{Jeon et~al.}{2015}]{Jeon_2015}
Jeon M.,  Bromm V.,  Pawlik A.~H.,   Milosavljevi\'{c} M.,  2015, \mn@doi
  [MNRAS] {10.1093/mnras/stv1353}, 452, 1152

\bibitem[\protect\citeauthoryear{Johnson \& Haardt}{Johnson \&
  Haardt}{2016}]{Johnson_Haardt_2016}
Johnson J.~L.,  Haardt F.,  2016, \mn@doi [PASA] {10.1017/pasa.2016.4}, 33,
  e007

\bibitem[\protect\citeauthoryear{Krumholz, Stone  \& Gardiner}{Krumholz
  et~al.}{2007}]{Krumholz_Stone_2007}
Krumholz M.~R.,  Stone J.~M.,   Gardiner T.~A.,  2007, \mn@doi [ApJ]
  {10.1086/522665}, 671, 518

\bibitem[\protect\citeauthoryear{Latif, Zaroubi  \& Spaans}{Latif
  et~al.}{2011}]{Latif_2011}
Latif M.~A.,  Zaroubi S.,   Spaans M.,  2011, \mn@doi [MNRAS]
  {10.1111/j.1365-2966.2010.17796.x}, 411, 1659

\bibitem[\protect\citeauthoryear{Laursen, Razoumov  \& Sommer-Larsen}{Laursen
  et~al.}{2009}]{Laursen_2009}
Laursen P.,  Razoumov A.~O.,   Sommer-Larsen J.,  2009, \mn@doi [ApJ]
  {10.1088/0004-637X/696/1/853}, 696, 853

\bibitem[\protect\citeauthoryear{Laursen, Sommer-Larsen  \& Razoumov}{Laursen
  et~al.}{2011}]{Laursen_2011}
Laursen P.,  Sommer-Larsen J.,   Razoumov A.~O.,  2011, \mn@doi [ApJ]
  {10.1088/0004-637X/728/1/52}, 728, 52

\bibitem[\protect\citeauthoryear{Loeb \& Furlanetto}{Loeb \&
  Furlanetto}{2013}]{Loeb_Furlanetto_2013}
Loeb A.,  Furlanetto S.~R.,  2013, {The First Galaxies in the Universe}.
Princeton Univ. Press, Princeton, NJ

\bibitem[\protect\citeauthoryear{Loeb \& Rybicki}{Loeb \&
  Rybicki}{1999}]{Loeb_Rybicki_1999}
Loeb A.,  Rybicki G.~B.,  1999, \mn@doi [ApJ] {10.1086/307844}, 524, 527

\bibitem[\protect\citeauthoryear{Madau \& Rees}{Madau \&
  Rees}{2000}]{Madau_Rees_2000}
Madau P.,  Rees M.~J.,  2000, \mn@doi [ApJ] {10.1086/312934}, 542, L69

\bibitem[\protect\citeauthoryear{Matthee, Sobral, Santos, R{\"{o}}ttgering,
  Darvish  \& Mobasher}{Matthee et~al.}{2015}]{Matthee_2015}
Matthee J.,  Sobral D.,  Santos S.,  R{\"{o}}ttgering H.,  Darvish B.,
  Mobasher B.,  2015, \mn@doi [MNRAS] {10.1093/mnras/stv947}, 451, 400

\bibitem[\protect\citeauthoryear{McKee \& Tan}{McKee \&
  Tan}{2008}]{McKee_Tan_2008}
McKee C.~F.,  Tan J.~C.,  2008, \mn@doi [ApJ] {10.1086/587434}, 681, 771

\bibitem[\protect\citeauthoryear{Mezzacappa \& Bruenn}{Mezzacappa \&
  Bruenn}{1993}]{Mezzacappa_1993}
Mezzacappa A.,  Bruenn S.~W.,  1993, \mn@doi [ApJ] {10.1086/172395}, 405, 669

\bibitem[\protect\citeauthoryear{Mihalas \& Auer}{Mihalas \&
  Auer}{2001}]{Mihalas_2001}
Mihalas D.,  Auer L.~H.,  2001, \mn@doi [J. Quant. Spectrosc. Radiative
  Transfer] {10.1016/S0022-4073(01)00013-9}, 71, 61

\bibitem[\protect\citeauthoryear{Mihalas \& Mihalas}{Mihalas \&
  Mihalas}{1984}]{Mihalas_book_1984}
Mihalas D.,  Mihalas B.~W.,  1984, {Foundations of radiation hydrodynamics}

\bibitem[\protect\citeauthoryear{Milosavljevi\'{c}, Bromm, Couch  \&
  Oh}{Milosavljevi\'{c} et~al.}{2009}]{Milosavljevic_2009}
Milosavljevi\'{c} M.,  Bromm V.,  Couch S.~M.,   Oh S.~P.,  2009, \mn@doi [ApJ]
  {10.1088/0004-637X/698/1/766}, 698, 766

\bibitem[\protect\citeauthoryear{Mortlock et~al.,}{Mortlock
  et~al.}{2011}]{Mortlock_2011}
Mortlock D.~J.,  et~al., 2011, \mn@doi [Nature] {10.1038/nature10159}, 474, 616

\bibitem[\protect\citeauthoryear{Neufeld}{Neufeld}{1991}]{Neufeld_1991}
Neufeld D.~A.,  1991, \mn@doi [ApJ] {10.1086/185983}, 370, L85

\bibitem[\protect\citeauthoryear{Norman, Reynolds, So, Harkness  \&
  Wise}{Norman et~al.}{2015}]{Norman_2015}
Norman M.~L.,  Reynolds D.~R.,  So G.~C.,  Harkness R.~P.,   Wise J.~H.,  2015,
  \mn@doi [ApJS] {10.1088/0067-0049/216/1/16}, 216, 16

\bibitem[\protect\citeauthoryear{Oesch et~al.,}{Oesch
  et~al.}{2015}]{Oesch_van_Dokkum_2015}
Oesch P.~A.,  et~al., 2015, \mn@doi [ApJ] {10.1088/2041-8205/804/2/L30}, 804,
  L30

\bibitem[\protect\citeauthoryear{Oesch et~al.,}{Oesch
  et~al.}{2016}]{Oesch_2016}
Oesch P.~A.,  et~al., 2016, \mn@doi [ApJ] {10.3847/0004-637X/819/2/129}, 819,
  129

\bibitem[\protect\citeauthoryear{Oh \& Haiman}{Oh \&
  Haiman}{2002}]{Oh_Haiman_2002}
Oh S.~P.,  Haiman Z.,  2002, \mn@doi [ApJ] {10.1086/339393}, 569, 558

\bibitem[\protect\citeauthoryear{{Osterbrock} \& {Ferland}}{{Osterbrock} \&
  {Ferland}}{2006}]{Osterbrock_book_2006}
{Osterbrock} D.~E.,  {Ferland} G.~J.,  2006, {Astrophysics of gaseous nebulae
  and active galactic nuclei, 2$^{\rm nd}$.~ed.}.
University Science Books, Sausalito, CA

\bibitem[\protect\citeauthoryear{Paardekooper, Khochfar  \& {Dalla
  Vecchia}}{Paardekooper et~al.}{2015}]{Paardekooper_2015}
Paardekooper J.-P.,  Khochfar S.,   {Dalla Vecchia} C.,  2015, \mn@doi [MNRAS]
  {10.1093/mnras/stv1114}, 451, 2544

\bibitem[\protect\citeauthoryear{Pacucci, Ferrara, Volonteri  \& Dubus}{Pacucci
  et~al.}{2015}]{Pacucci_MBH_Spectra_2015}
Pacucci F.,  Ferrara A.,  Volonteri M.,   Dubus G.,  2015, \mn@doi [MNRAS]
  {10.1093/mnras/stv2196}, 454, 3771

\bibitem[\protect\citeauthoryear{Pacucci, Ferrara, Grazian, Fiore  \&
  Giallongo}{Pacucci et~al.}{2016}]{Pacucci_DCBH_2016}
Pacucci F.,  Ferrara A.,  Grazian A.,  Fiore F.,   Giallongo E.,  2016, \mn@doi
  [MNRAS] {10.1093/mnras/stw725}, 459, 1432

\bibitem[\protect\citeauthoryear{Pallottini et~al.,}{Pallottini
  et~al.}{2015}]{Pallottini_2015}
Pallottini A.,  et~al., 2015, \mn@doi [MNRAS] {10.1093/mnras/stv1795}, 453,
  2465

\bibitem[\protect\citeauthoryear{Partridge \& Peebles}{Partridge \&
  Peebles}{1967}]{Partridge_Peebles_1967}
Partridge R.~B.,  Peebles P. J.~E.,  1967, \mn@doi [ApJ] {10.1086/149079}, 147,
  868

\bibitem[\protect\citeauthoryear{Pawlik \& Schaye}{Pawlik \&
  Schaye}{2011}]{Pawlik_2011}
Pawlik A.~H.,  Schaye J.,  2011, \mn@doi [MNRAS]
  {10.1111/j.1365-2966.2010.18032.x}, 412, 1943

\bibitem[\protect\citeauthoryear{Pawlik, Milosavljevi\'{c}  \& Bromm}{Pawlik
  et~al.}{2013}]{Pawlik_2013}
Pawlik A.~H.,  Milosavljevi\'{c} M.,   Bromm V.,  2013, \mn@doi [ApJ]
  {10.1088/0004-637X/767/1/59}, 767, 59

\bibitem[\protect\citeauthoryear{Raiter, Schaerer  \& Fosbury}{Raiter
  et~al.}{2010}]{Raiter_2010}
Raiter A.,  Schaerer D.,   Fosbury R. A.~E.,  2010, \mn@doi [A{\&}A]
  {10.1051/0004-6361/201015236}, 523, A64

\bibitem[\protect\citeauthoryear{Rees \& Ostriker}{Rees \&
  Ostriker}{1977}]{Rees_Ostriker_1977}
Rees M.~J.,  Ostriker J.~P.,  1977, MNRAS, 179, 541

\bibitem[\protect\citeauthoryear{Roth \& Kasen}{Roth \&
  Kasen}{2015}]{Roth_2015}
Roth N.,  Kasen D.,  2015, \mn@doi [ApJS] {10.1088/0067-0049/217/1/9}, 217, 9

\bibitem[\protect\citeauthoryear{Schaerer}{Schaerer}{2002}]{Schaerer_2002}
Schaerer D.,  2002, \mn@doi [A\&A] {10.1051/0004-6361:20011619}, 382, 28

\bibitem[\protect\citeauthoryear{Shull \& van Steenberg}{Shull \& van
  Steenberg}{1985}]{Shull_1985}
Shull J.~M.,  van Steenberg M.~E.,  1985, \mn@doi [ApJ] {10.1086/163605}, 298,
  268

\bibitem[\protect\citeauthoryear{Smidt, Wiggins  \& Johnson}{Smidt
  et~al.}{2016}]{Smidt_CR7_2016}
Smidt J.,  Wiggins B.~K.,   Johnson J.~L.,  2016, ApJ, preprint
  (\href{http://arxiv.org/abs/1603.00888}{arXiv:1603.00888})

\bibitem[\protect\citeauthoryear{Smith, Safranek-Shrader, Bromm  \&
  Milosavljevi\'{c}}{Smith et~al.}{2015}]{Smith_2015}
Smith A.,  Safranek-Shrader C.,  Bromm V.,   Milosavljevi\'{c} M.,  2015,
  \mn@doi [MNRAS] {10.1093/mnras/stv565}, 449, 4336

\bibitem[\protect\citeauthoryear{Smith, Bromm  \& Loeb}{Smith
  et~al.}{2016}]{Smith_CR7_2016}
Smith A.,  Bromm V.,   Loeb A.,  2016, \mn@doi [MNRAS] {10.1093/mnras/stw1129},
  460, 3143

\bibitem[\protect\citeauthoryear{So, Norman, Reynolds  \& Wise}{So
  et~al.}{2014}]{So_2015}
So G.~C.,  Norman M.~L.,  Reynolds D.~R.,   Wise J.~H.,  2014, \mn@doi [ApJ]
  {10.1088/0004-637X/789/2/149}, 789, 149

\bibitem[\protect\citeauthoryear{Sobral, Matthee, Darvish, Schaerer, Mobasher,
  R{\"{o}}ttgering, Santos  \& Hemmati}{Sobral et~al.}{2015}]{Sobral_2015}
Sobral D.,  Matthee J.,  Darvish B.,  Schaerer D.,  Mobasher B.,
  R{\"{o}}ttgering H. J.~A.,  Santos S.,   Hemmati S.,  2015, \mn@doi [ApJ]
  {10.1088/0004-637X/808/2/139}, 808, 139

\bibitem[\protect\citeauthoryear{Somerville \& Dav{\'{e}}}{Somerville \&
  Dav{\'{e}}}{2015}]{Somerville_Dave_2015}
Somerville R.~S.,  Dav{\'{e}} R.,  2015, \mn@doi [ARA\&A]
  {10.1146/annurev-astro-082812-140951}, 53, 51

\bibitem[\protect\citeauthoryear{Spaans \& Silk}{Spaans \&
  Silk}{2006}]{Spaans_Silk_2006}
Spaans M.,  Silk J.,  2006, \mn@doi [ApJ] {10.1086/508444}, 652, 902

\bibitem[\protect\citeauthoryear{Stacy, Greif  \& Bromm}{Stacy
  et~al.}{2012}]{Stacy_2012}
Stacy A.,  Greif T.~H.,   Bromm V.,  2012, \mn@doi [MNRAS]
  {10.1111/j.1365-2966.2012.20605.x}, 422, 290

\bibitem[\protect\citeauthoryear{Stark et~al.,}{Stark
  et~al.}{2015}]{Stark_2015}
Stark D.~P.,  et~al., 2015, \mn@doi [MNRAS] {10.1093/mnras/stv688}, 450, 1846

\bibitem[\protect\citeauthoryear{Struve}{Struve}{1942}]{Struve_1942}
Struve O.,  1942, \mn@doi [ApJ] {10.1086/144380}, 95, 134

\bibitem[\protect\citeauthoryear{Tasitsiomi}{Tasitsiomi}{2006}]{Tasitsiomi_2006}
Tasitsiomi A.,  2006, \mn@doi [ApJ] {10.1086/504460}, 645, 792

\bibitem[\protect\citeauthoryear{Tsang \& Milosavljevi\'{c}}{Tsang \&
  Milosavljevi\'{c}}{2015}]{Tsang_2015}
Tsang B. T.-H.,  Milosavljevi\'{c} M.,  2015, \mn@doi [MNRAS]
  {10.1093/mnras/stv1707}, 453, 1108

\bibitem[\protect\citeauthoryear{{Vanden Berk} et~al.,}{{Vanden Berk}
  et~al.}{2001}]{Vanden_Berk_2001}
{Vanden Berk} D.~E.,  et~al., 2001, \mn@doi [AJ] {10.1086/321167}, 122, 549

\bibitem[\protect\citeauthoryear{Verhamme, Schaerer  \& Maselli}{Verhamme
  et~al.}{2006}]{Verhamme_2006}
Verhamme A.,  Schaerer D.,   Maselli A.,  2006, \mn@doi [A\&A]
  {10.1051/0004-6361:20065554}, 460, 397

\bibitem[\protect\citeauthoryear{Verhamme, Dubois, Blaizot, Garel, Bacon,
  Devriendt, Guiderdoni  \& Slyz}{Verhamme et~al.}{2012}]{Verhamme_2012}
Verhamme A.,  Dubois Y.,  Blaizot J.,  Garel T.,  Bacon R.,  Devriendt J.,
  Guiderdoni B.,   Slyz A.,  2012, \mn@doi [A\&A]
  {10.1051/0004-6361/201218783}, 546, A111

\bibitem[\protect\citeauthoryear{Visbal, Haiman  \& Bryan}{Visbal
  et~al.}{2016}]{Visbal_CR7_2016}
Visbal E.,  Haiman Z.,   Bryan G.~L.,  2016, \mn@doi [MNRAS]
  {10.1093/mnrasl/slw071}, 460, L59

\bibitem[\protect\citeauthoryear{Wise \& Cen}{Wise \& Cen}{2009}]{Wise_2009}
Wise J.~H.,  Cen R.,  2009, \mn@doi [ApJ] {10.1088/0004-637X/693/1/984}, 693,
  984

\bibitem[\protect\citeauthoryear{Wise, Abel, Turk, Norman  \& Smith}{Wise
  et~al.}{2012}]{Wise_2012}
Wise J.~H.,  Abel T.,  Turk M.~J.,  Norman M.~L.,   Smith B.~D.,  2012, \mn@doi
  [MNRAS] {10.1111/j.1365-2966.2012.21809.x}, 427, 311

\bibitem[\protect\citeauthoryear{Wu et~al.,}{Wu et~al.}{2015}]{Wu_2015}
Wu X.-B.,  et~al., 2015, \mn@doi [Nature] {10.1038/nature14241}, 518, 512

\bibitem[\protect\citeauthoryear{Xu, Wise, Norman, Ahn  \& O'Shea}{Xu
  et~al.}{2016}]{Xu_2016}
Xu H.,  Wise J.~H.,  Norman M.~L.,  Ahn K.,   O'Shea B.~W.,  2016, ApJ,
  preprint (\href{http://arxiv.org/abs/1604.07842}{arXiv:1604.07842})

\bibitem[\protect\citeauthoryear{Yajima, Choi  \& Nagamine}{Yajima
  et~al.}{2011}]{Yajima_2011}
Yajima H.,  Choi J.-H.,   Nagamine K.,  2011, \mn@doi [MNRAS]
  {10.1111/j.1365-2966.2010.17920.x}, 412, 411

\bibitem[\protect\citeauthoryear{Zanstra}{Zanstra}{1934}]{Zanstra_1934}
Zanstra H.,  1934, MNRAS, 95, 84

\bibitem[\protect\citeauthoryear{Zheng \& Miralda-Escud\'{e}}{Zheng \&
  Miralda-Escud\'{e}}{2002}]{Zheng_2002}
Zheng Z.,  Miralda-Escud\'{e} J.,  2002, \mn@doi [ApJ] {10.1086/342400}, 578,
  33

\bibitem[\protect\citeauthoryear{Zheng \& Wallace}{Zheng \&
  Wallace}{2014}]{Zheng_Wallace_2014}
Zheng Z.,  Wallace J.,  2014, \mn@doi [ApJ] {10.1088/0004-637X/794/2/116}, 794,
  116

\bibitem[\protect\citeauthoryear{Zitrin et~al.,}{Zitrin
  et~al.}{2015}]{Zitrin_2015}
Zitrin A.,  et~al., 2015, \mn@doi [ApJ] {10.1088/2041-8205/810/1/L12}, 810, L12

\bibitem[\protect\citeauthoryear{{von Neumann} \& Richtmyer}{{von Neumann} \&
  Richtmyer}{1950}]{Von_Neumann_1950}
{von Neumann} J.,  Richtmyer R.~D.,  1950, \mn@doi [J. Appl. Phys.]
  {10.1063/1.1699639}, 21, 232

\makeatother
\end{thebibliography}

\bsp 
\label{lastpage}
\end{document}